\documentclass[epj]{svjour}
\usepackage{amssymb}
\usepackage{url}
\usepackage{graphics}
\usepackage[normalem]{ulem}   
\usepackage{color}  
\usepackage{graphics}
\usepackage{graphicx}
\usepackage{amssymb}
\usepackage{amsfonts}
\usepackage{amsmath}
\usepackage{amsbsy}
\usepackage[numbers,sort&compress]{natbib}
\usepackage{comment}
\usepackage{color}
\usepackage{subfigure}

\begin{document}

\title{Scrambling and quantum feedback in a nanomechanical system\footnote{Please insert title footnote here}}
\author{A. K. Singh\inst{1}, Kushagra Sachan\inst{1} \and L.~Chotorlishvili\inst{2}, Vipin V.\inst{1} \and Sunil~K.~Mishra\inst{1} 
}                     
%
%
\institute{Department of Physics, Indian Institute of Technology (Banaras Hindu University), Varanasi - 221005, India \and Faculty of Mathematics and Natural Sciences, Tbilisi State University, Chavchavadze av.3, 0128 Tbilisi, Georgia}
\date{Received: date / Revised version: date}
%
\abstract{
The question of how swiftly entanglement spreads over a system has attracted vital interest. In this regard, the out-of-time ordered correlator (OTOC) is a quantitative measure of the entanglement spreading process. Particular interest concerns the propagation of quantum correlations in the lattice systems, {\it e.g.},~spin chains. In a seminal paper D.~A.~Roberts, D.~Stanford and L.~Susskind, J.~High Energy Phys.~{\bf 03}, 051, (2015) the concept of the OTOC's radius was introduced. The radius of the OTOC defines the front line reached by the spread of entanglement. Beyond this radius operators commute. In the present work, we propose a model of two nanomechanical systems coupled with two Nitrogen-vacancy (NV) center spins. Oscillators are coupled to each other directly while NV spins are not. Therefore, the correlation between the NV spins may arise only through the quantum feedback exerted from the first NV spin to the first oscillator and transferred from the first oscillator to the second oscillator via the direct coupling. Thus nonzero OTOC between NV spins quantifies the strength of the quantum feedback. We show that NV spins cannot exert quantum feedback on classical nonlinear oscillators. We also discuss inherently quantum case with a linear quantum harmonic oscillator indirectly coupling the two spins and verify that in the classical limit of the oscillator, the OTOC vanishes. 
%
} 
\maketitle
\section{Introduction}
\label{intro}
A sudden quench of parameters on the quantum state of a system leads to reshuffling of quantum information, such as the entanglement stored in a many-body correlated initial quantum state\cite{Heyl2018,Heyl2018a,Heyl2013,Vosk2014,Eisert2015,Ponte2015,Azimi2014,Azimi2016,PhysRevA.103.052216}, during the subsequent time evolution. An important question is the swiftness of the spreading of the quantum entanglement.  The maximum rate at which correlations buildup in the quantum system is limited by the
Lieb-Robinson bound \cite{liebrobinson}, while a quantitative criterion is provided by the out-of-time-ordered correlation (OTOC) of the operators in question. The concept of the OTOC was introduced by Larkin and Ovchinnikov\cite{larkin}, and since then, OTOC has been seen as a diagnostic tool of quantum chaos. Interest in the delocalization of quantum information (i.e., the scrambling of quantum entanglement) was renewed only recently, see Refs.~\cite{Maldacena2016,Roberts2015,Iyoda2018,Chapman2018,Swingle2017,Klug2018,Campo2017,Campisi2017,Grozdanov2018,Patel2017,Khemani2018,Rakovszky2018,Syzranov2018,Hosur2016,Halpern2017} and references therein. In the present work, we show that OTOC can be exploited as a quantifier of quantum feedback. In particular, we propose a model of two nanomechanical systems (nonlinear oscillators) coupled with two NV spins. We prove that entanglement between two NV centers can spread only if NV centers are connected through the quantum channel. In the semi-classical and classical channel limit, entanglement decays to zero.

Let us consider two unitary operators $\hat{V}$ and $\hat{W}$, describing the local perturbations to the system, and their unitary time evolution under a Hamiltonian $\hat{H}$, which we will specify as, ${\hat{W}\left(t\right)=\exp (i\hat{H}t)\,\hat{W}\exp(-i\hat{H}t)}$. Here we measure time such that $\hbar=1$. Then the OTOC is defined as

\begin{equation}\label{OTOC1}
C\left(t\right)= \frac{1}{2}\left\langle\left[\hat{W}(t),\hat{V}\right]^{\dag}\left[\hat{W}(t),\hat{V}\right]\right\rangle\,,
\end{equation}

or in an equivalent form as 

\begin{equation}\label{_but_equivalent}
C\left(t\right)=1-\operatorname{Re} \mathcal{F}\left(t\right),
\end{equation}
where $\mathcal{F}\left(t\right)=\left\langle \hat{W}(t)^{\dagger}\hat{V}^{\dagger}\hat{W}(t)\hat{V}\right\rangle$. Here, parentheses $\langle\ldots\rangle$, if not otherwise specified, denote either the quantum mechanical ground state average $\langle\ldots\rangle$=$\langle\psi|\ldots|\psi\rangle$, or the finite temperature thermal average $\langle\ldots\rangle=Z^{-1}{\rm Tr}\left(e^{-\beta \hat{H}}\ldots\right)$ with inverse temperature ${\beta=1/T}$ and the Boltzmann constant scaled to ${k_B=1}$. At the initial moment of time, as follows from the definition, the OTOC is zero ${C(0)=0}$, provided ${[\hat{W},\hat{V}]=0}$.

OTOCs demonstrate several interesting physical features in the integrable and nonintegrable systems. For example, magnetization OTOCs ($V=W=$Magnetization) point to dynamical phase transitions\cite{Heyl2018a}. Disorder slows the growth of  $C\left(t\right)$ in time, and therefore, scrambling can be used to identify the many-body localization phase\cite{Swingle2017}. Scrambling itself is nothing other than the zero velocity Lieb Robinson bound\cite{Hamza2012}.
\begin{equation}\label{OTOC2}
C\left(t\right)=||[W(t),V]||={\rm min}\left(|t|,1\right)e^{-\eta d(\hat{W},\hat{V})},.
\end{equation}

Here, ${\eta=\mathrm{const}}$, $d(\hat{W}, \hat{V})$ is the distance between operators on the lattice (i.e.~spin operators $\exp (\hat{S}_{n}^{z})$ of a spin chain), and the Frobenius norm of the unitary operator $\hat{A}$ is defined as $\|\hat{A}\|=\text{Tr}(\hat{A}^{\dag}\hat{A})$.

Rather interesting and enlightening is a semi-classical interpretation of scrambling. For canonical momentum and coordinate operators ${V\equiv P}$, ${W(t)\equiv q(t)}$ one deduces  the equation, valid on the short time scale \cite{Maldacena2016},\\ $C\left(t\right)=\hbar^{2}\exp\left(2\lambda_{L}t\right)$.  The scrambling time is specified in terms of the classical Lyapunov exponent $\lambda_{L}$ and is equal to the Ehrenfest time $\tau\approx\frac{1}{\lambda_{L}}\ln1/\hbar$. On the other hand, a purely quantum analysis shows that the radius of the operator linearly increases in time, independent of whether the quantum system is integrable or chaotic \cite{Roberts2015}. This means that in the qubit system (for example, a Heisenberg spin chain), the time required for the formation of correlations between initially commuting operators $\left[\sigma_{n}^{j},\sigma_{m}^{k}\right]=2i\delta_{nm}\epsilon^{jkl}\sigma^{l}$ will increase linearly with the distance ${d=|n-m|}$ between them, and hence $\left[\sigma_{n}^{j}\left(t\right),\sigma_{m}^{k}\right]\neq 0$, for ${n\neq m}$. Note that customarily scrambling is an irreversible process after entanglement is spread across the system, it cannot be unscrambled\cite{Campisi2017}.

Indeed OTOC can be interpreted in terms of two wave functions time evolved in a different manner. Let $|\psi(0)\rangle$ be the initial pure state wave function which is time evolved in following steps: first it is perturbed at ${t=0}$ with a local unitary operator $\hat{V}$, then evolved forward under the unitary evolution operator ${\hat{U}=\exp (-i\hat{H}t)}$ until ${t=\tau}$, it is then perturbed at ${t=\tau}$ with a local unitary operator $\hat{W}$, and evolved backward from ${t=\tau}$ to ${t=2\tau}$ under $\hat{U^{\dagger}}$. Hence the time evolved wave function is $|\psi(2\tau)\rangle =\hat{U^{\dagger}}\hat{W}\hat{U}\hat{V}|\psi(0)\rangle=\hat{W}(t)\hat{V}|\psi(0)\rangle$. For the second wave function the order of the applied perturbations is permuted, i.e.~first $\hat{W}$ at ${t=\tau}$ and then $\hat{V}$ at ${t=2\tau}$. Therefore the second wavefunction is $|\phi(2\tau)\rangle=\hat{V}\hat{U^{\dagger}}\hat{W}\hat{U}|\psi(0)\rangle=\hat{V}\hat{W}(t)|\psi(0)\rangle$ and their overlap is equivalent to the OTOC $\mathcal{F}(t)=\langle\phi(t)|\psi(t)\rangle$. What breaks the time inversion symmetry for the OTOC is the permuted sequence of operators $\hat{W}$ and $\hat{V}$. Concerning the interplay between OTOC and dynamical phase transitions, we admit a recent experimental work \cite{hainaut2018experimental}. Analyzing the behavior of the chaotic quantum system, authors experimentally observed the sudden change in the system’s memory behavior.

Recently interest is focused on the hybrid quantum-classical nanoelectromechanical systems (NEMS)
\cite{Naik2009,Connell2010,Alegre2011,Stannigel2010,Safavi-Naeini2011,Camerer2011,Eichenfield2009,Safavi-Naeini2012,Brahms2012,Nunnenkamp2012,Khalili2012,Meaney2011,Atalaya2011,Rabl2010,chotorlishvili2011thermal,Prants2011,Ludwig2010,Schmidt2010,Karabalin2009,Chotorlishvili_2011,Shevchenko2012,Liu2010,Shevchenko2010,Zueco2009,Cohen2013,Rabl2009,Zhou2010,Chotorlishvili2013}. Typically a NEMS consists of two parts: quantum NV center and classical cantilever (in what follows oscillator).
Therefore, NEMS may manifest binary quantum-classical features. A spin of the NV center couples with a cantilever through the magnetic tip attached to a cantilever. An oscillator performs the classical oscillations if not cooled down to the cryogenic temperatures. We are interested in the question if classical coupled nonlinear oscillators may transfer the quantum correlations. In other words, we aim to explore the problem of scrambling quantum information through the classical channel.  To answer this question, we study a system of two strongly coupled nonlinear NEMS.
\begin{figure}[!t]
    \centering
    \includegraphics[width=.5\textwidth,height=3.5in]{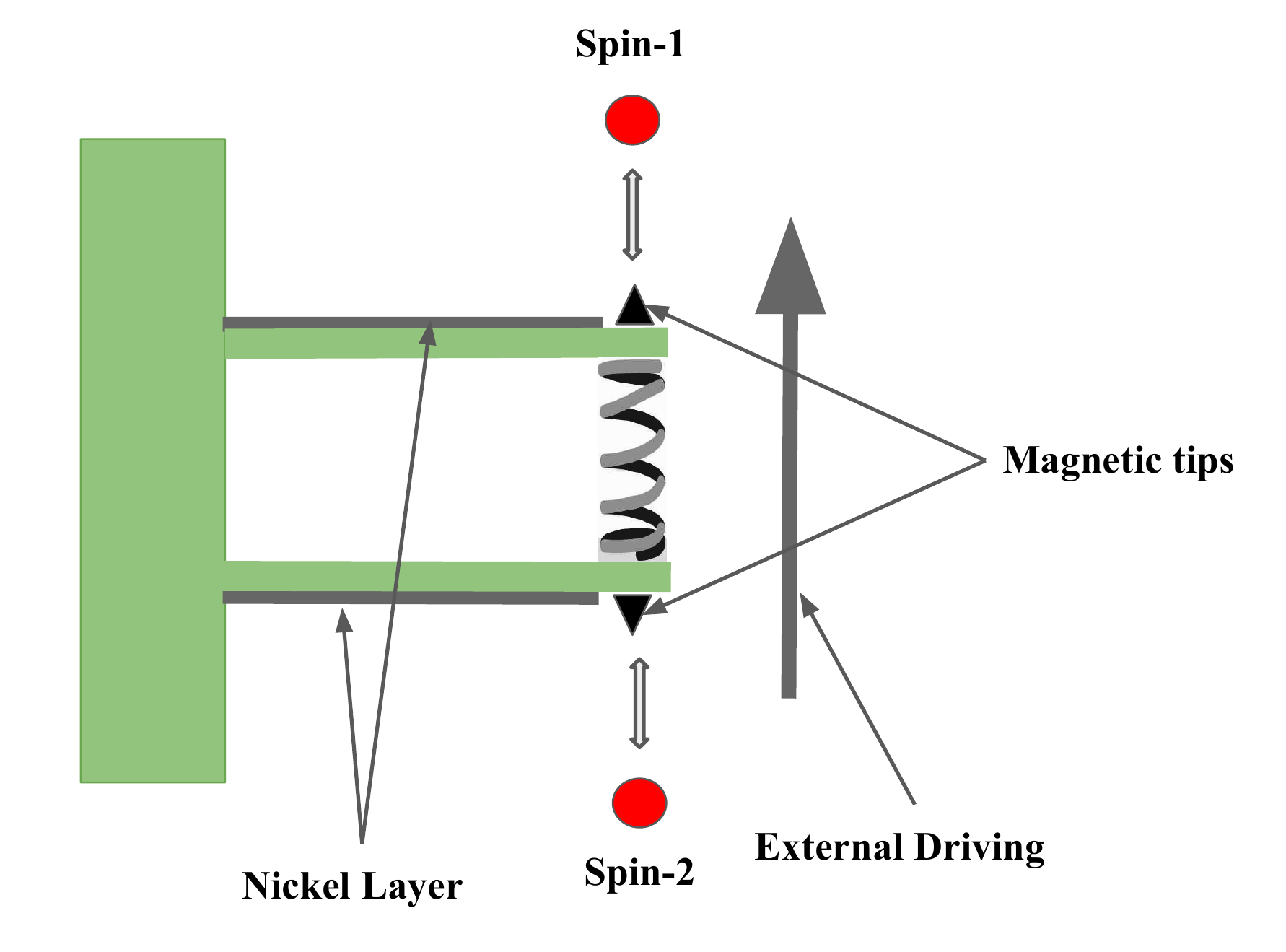}
    \caption{(Color online) Schematics of two NV spins coupled via coupled oscillators.  Oscillators are coupled to each other directly while NV spins are not. }
    \label{fig1}
\end{figure}
The devices consist of three layers of gallium arsenide (GaAs) the 100 nm \textbf{n}-doped layer, the 50 nm insulating layer, and the \textbf{p} doped 50 nm layer. For more details about the system we refer to the work \cite{PhysRevB.79.165309}.
We assume that each of the oscillators is coupled to the spin of NV center and consider OTOC as a measure of the quantum feedback. In this manuscript we will first discuss the model in section \ref{sec_model} and the scheme for numerical solution of the model. Afterwards, in section \ref{results} we will discuss the analytical solution of the model in the absence of quantum feedback followed by  a discussion on the numerical calculation of the system in various regimes. In section \ref{inherent_quantum} we will discuss a system of two spins coupled indirectly with a quantum linear oscillator. Finally we will conclude in section \ref{conclusion}. 
\section{Model}\label{sec_model}
The schematics of the system in question is shown in Fig. \ref{fig1}.
\begin{figure*}[!t]
 \includegraphics[width=0.51\textwidth,height=2in]{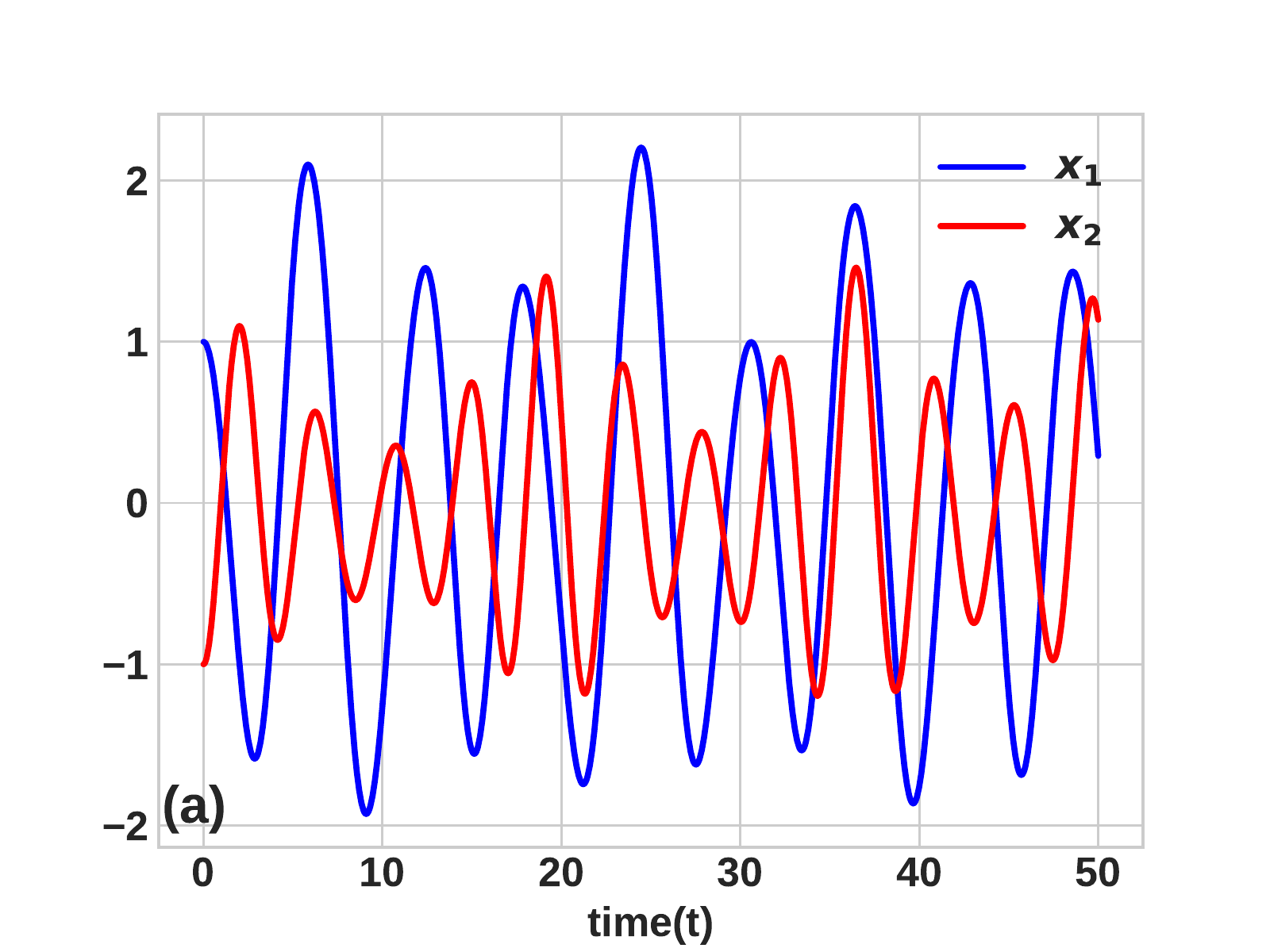}\ \includegraphics[width=0.51\textwidth,height=2in]{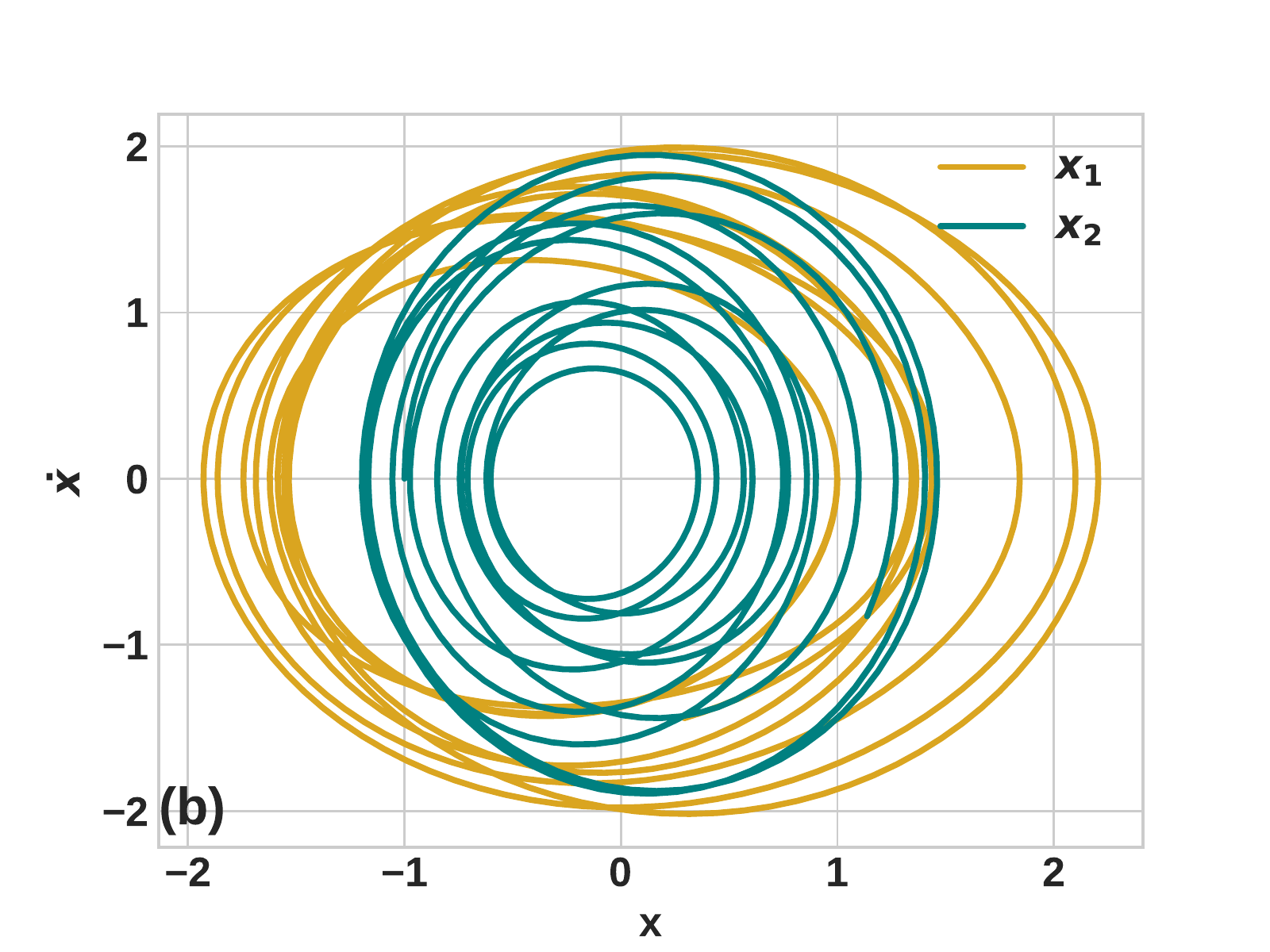}\\
 \includegraphics[width=0.51\textwidth,height=2in]{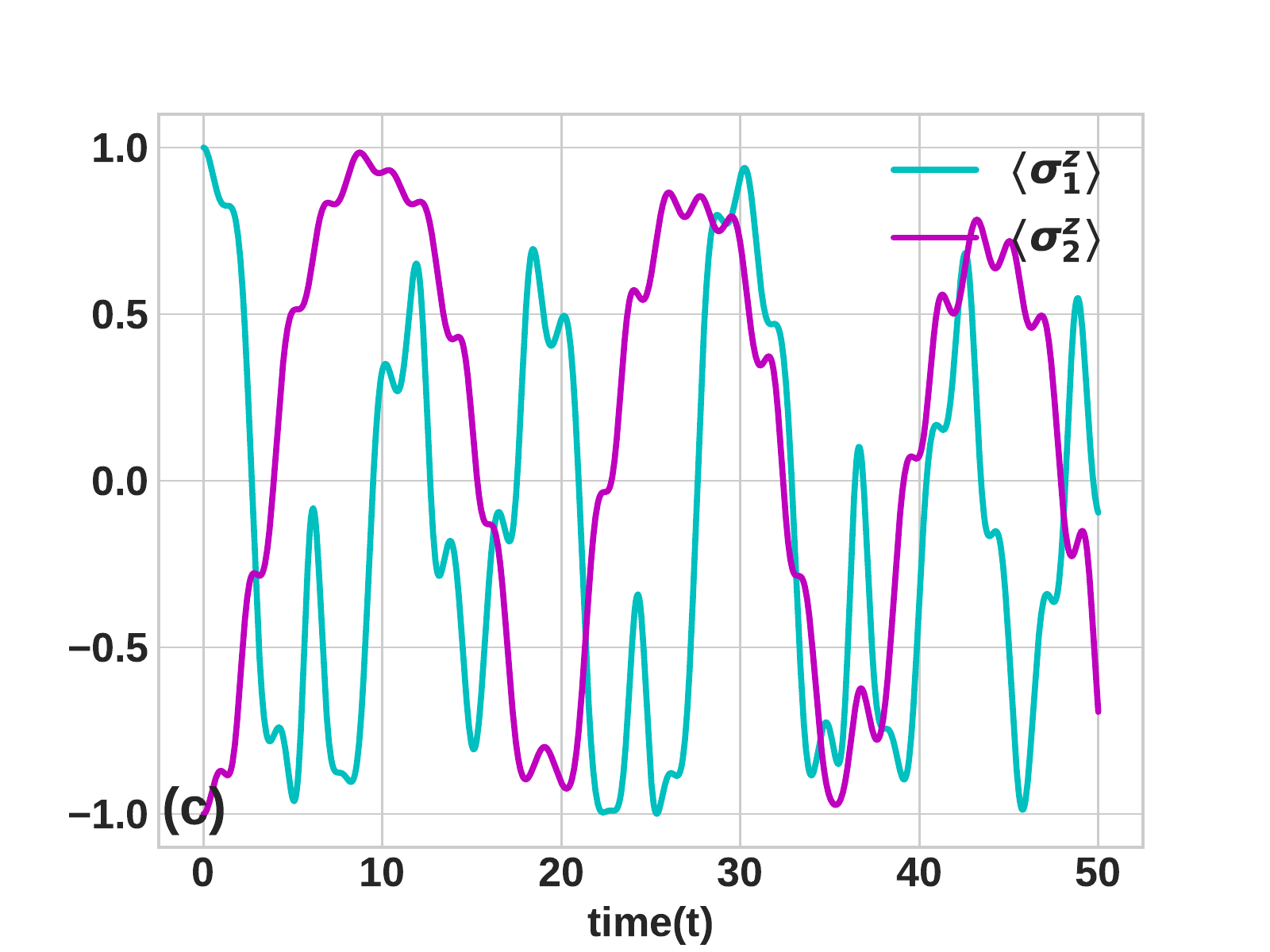}\ \includegraphics[width=0.51\textwidth,height=2in]{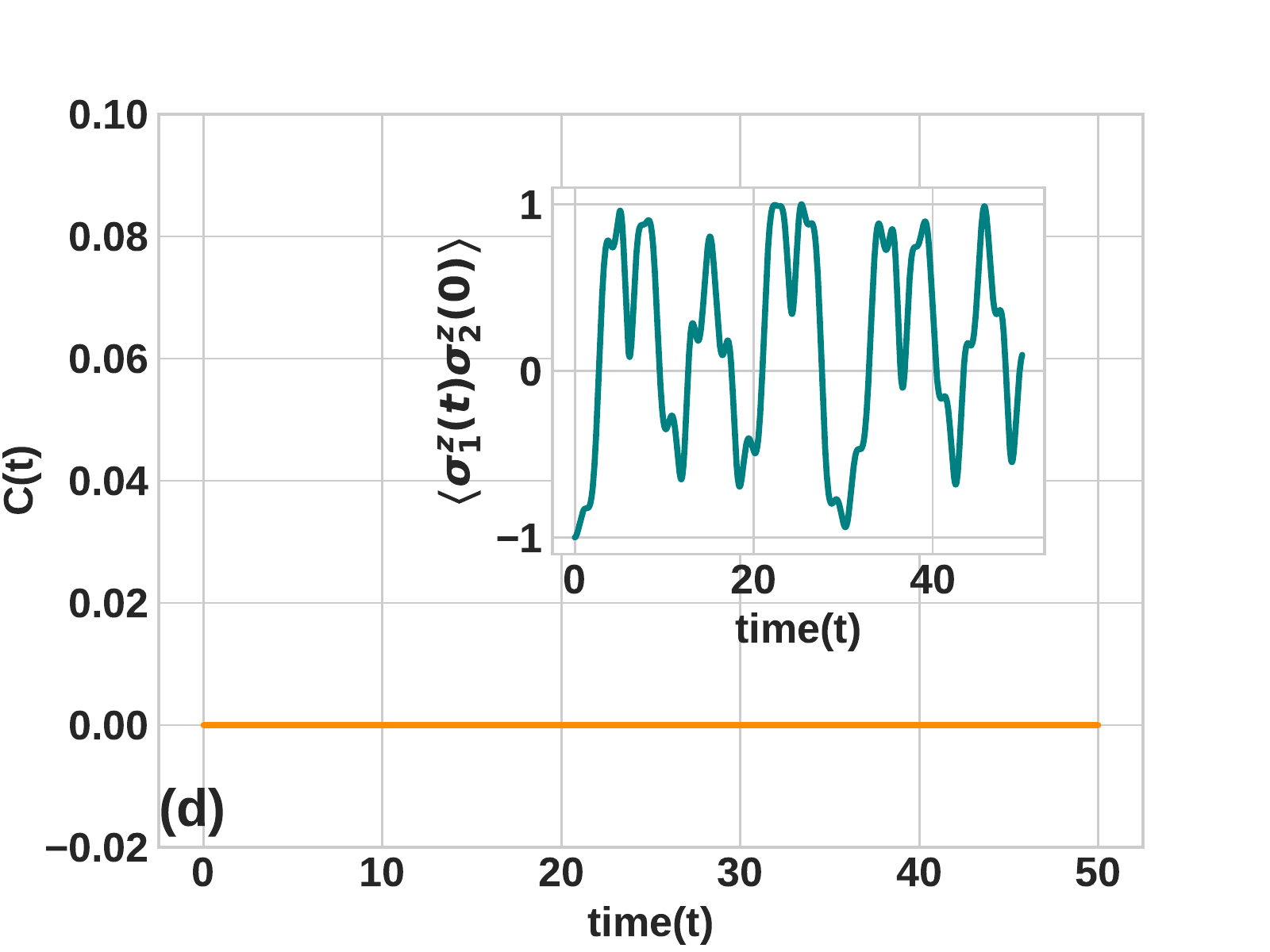}
\caption{Autonomous linear oscillators and weak connectivity regime: Position and phase space plots   in (a) and (b), and spin dynamics and OTOC in (c) and (d). Inset in (d) shows two-point time ordered correlation.  The values of the parameters are $\omega_{0}=1.5$, $\omega_{1}=1.0$, $\omega_{2}=1.5$, $F_{1}=F_{1}=0$, $\xi=0$, $\gamma=0$, $g=1$, $K=0.1$, $\alpha=\pi/3$. }
\label{regular}
 \end{figure*}
The two Oscillators are coupled to each other directly. The strength of the coupling between the oscillators depends on the coupling constant and eigenfrequencies of the oscillators. We quantify this coupling strength through the ``connectivity".  The first NV spin we coupled to the first oscillator and the second NV spin to the second oscillator. On the other hand, spins are not coupled directly and the correlation between NV spins may arise only through the quantum feedback exerted from the first NV spin to the first oscillator and transferred from the first oscillator to the second oscillator via the direct coupling. The Hamiltonian of the two NV spins coupled to the nonlinear oscillators reads: \cite{PhysRevB.101.104311}
\begin{eqnarray}\label{two NV spins}
&&\hat H_s=\frac{1}{2}\omega_0\left(\hat\sigma^z_1+\hat\sigma^z_2\right)+gx_1(t)\hat S^z_1+gx_2(t)\hat S^z_2.
\end{eqnarray}
Here $\omega_0=(\omega_R^2+\delta^2)^{1/2}$, $\omega_R$ is the Rabi frequency, and $\delta$ is the detuning between the microwave frequency and the intrinsic frequency of the NV spin. The operator $\hat S^z$ in the eigenbasis of the
NV center has the form $\hat S_{1,2}^z=\frac{1}{2}(\cos\alpha~\hat\sigma^z_{1,2}+\sin\alpha~(\hat\sigma_{1,2}^++\hat\sigma_{1,2}^-))$, where  $\hat\sigma^{\pm}=\frac{1}{2}(\hat\sigma_x\pm i\hat\sigma_y)$ and $\tan\alpha=-\omega_R/\delta$ and $g$ is interaction constant between the oscillators and the spins. A magnetic tip is attached to the end of the nanomechanical resonators. During the oscillation performed by the resonators, the distance between the NV spin and the magnetic tip $x$ changes. Oscillations produce a time-varying magnetic field, and therefore, the NV spin senses the motion of the magnetized resonator tip.   For more details about the origin of the coupling between resonator and NV spin, we refer to the work \cite{Rabl2009}. 
\begin{figure*}[!t]
 \includegraphics[width=0.51\textwidth,height=2in]{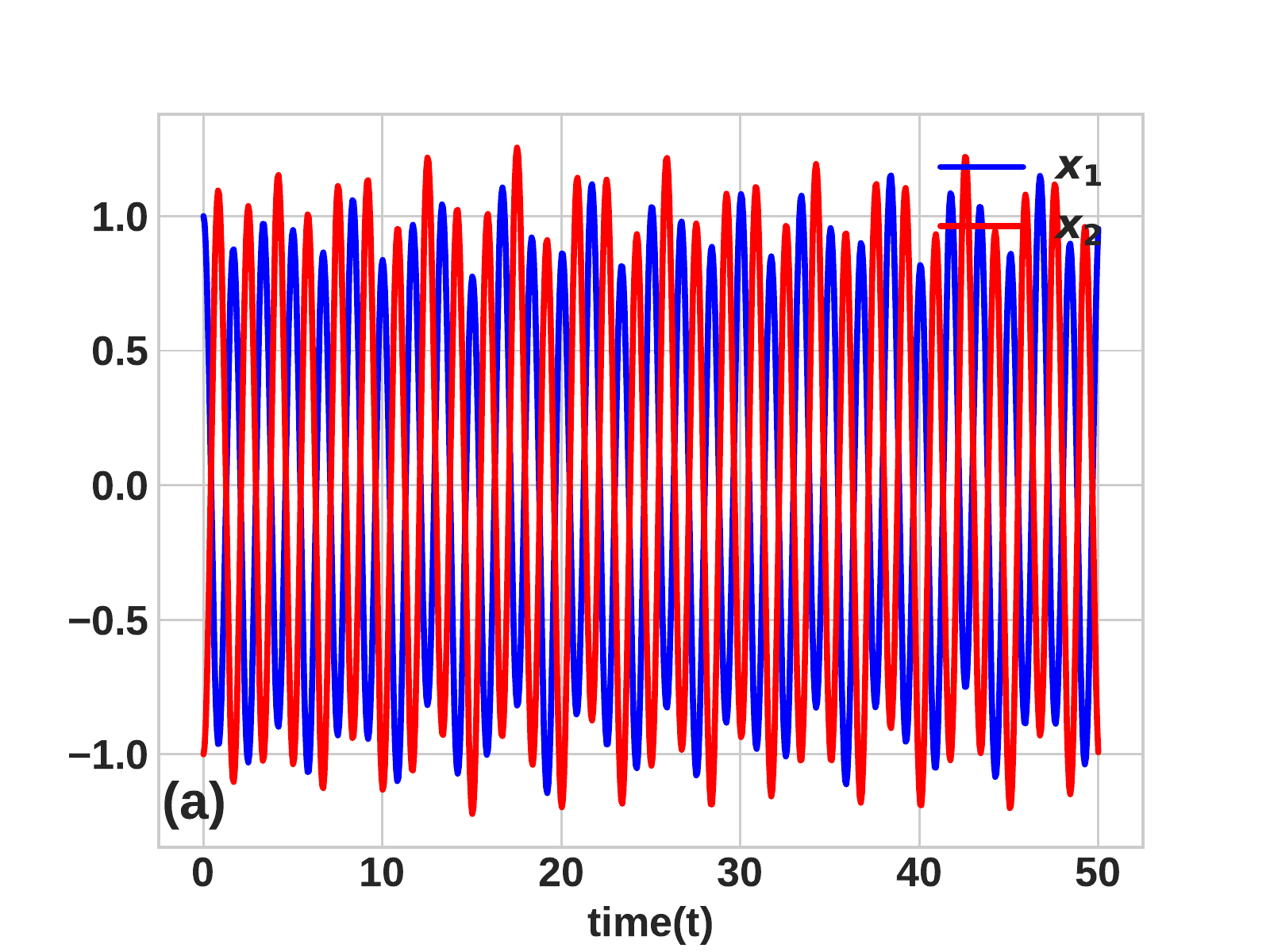}\ \includegraphics[width=0.51\textwidth,height=2in]{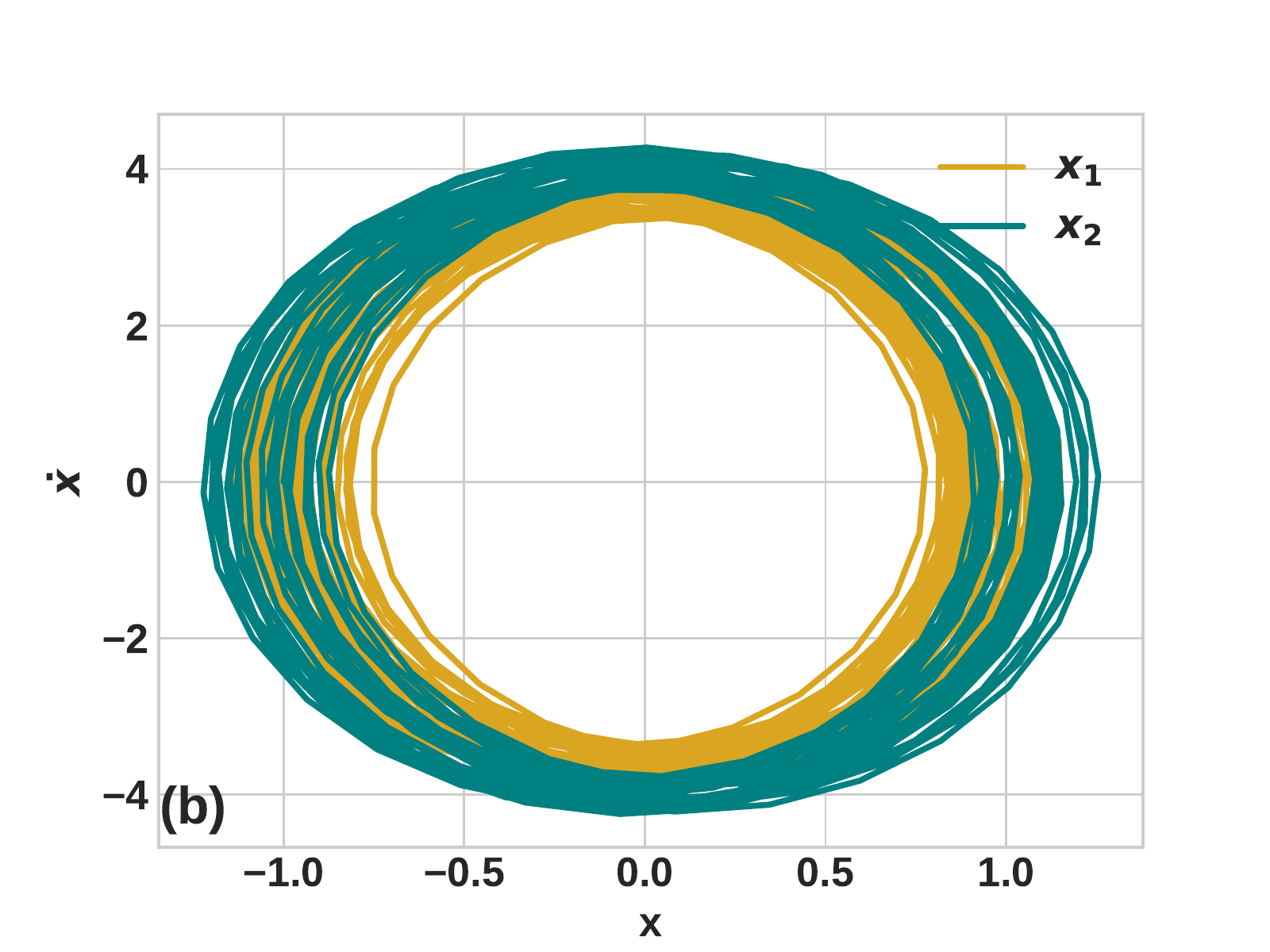}\\
 \includegraphics[width=0.51\textwidth,height=2in]{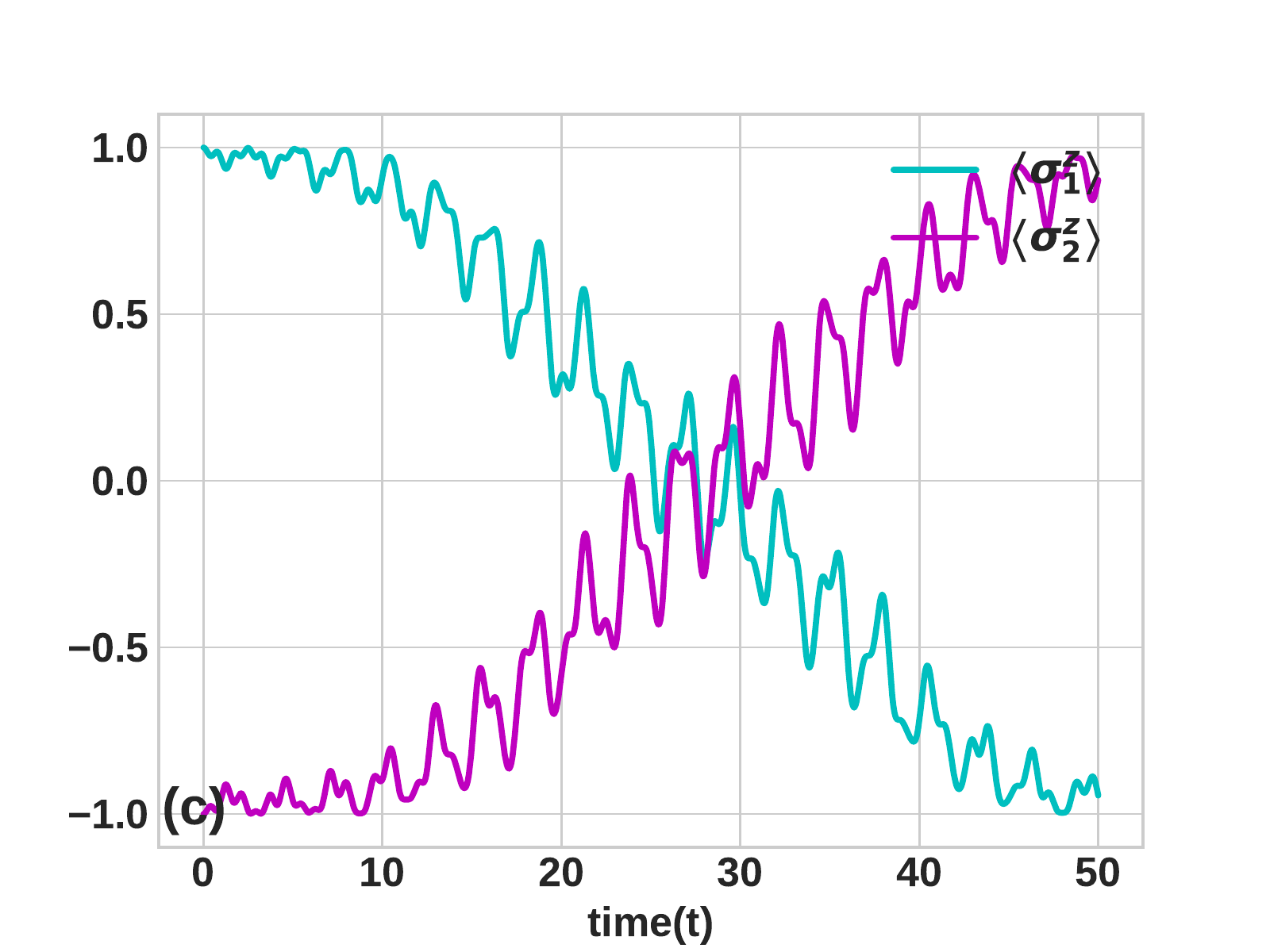}\ \includegraphics[width=0.51\textwidth,height=2in]{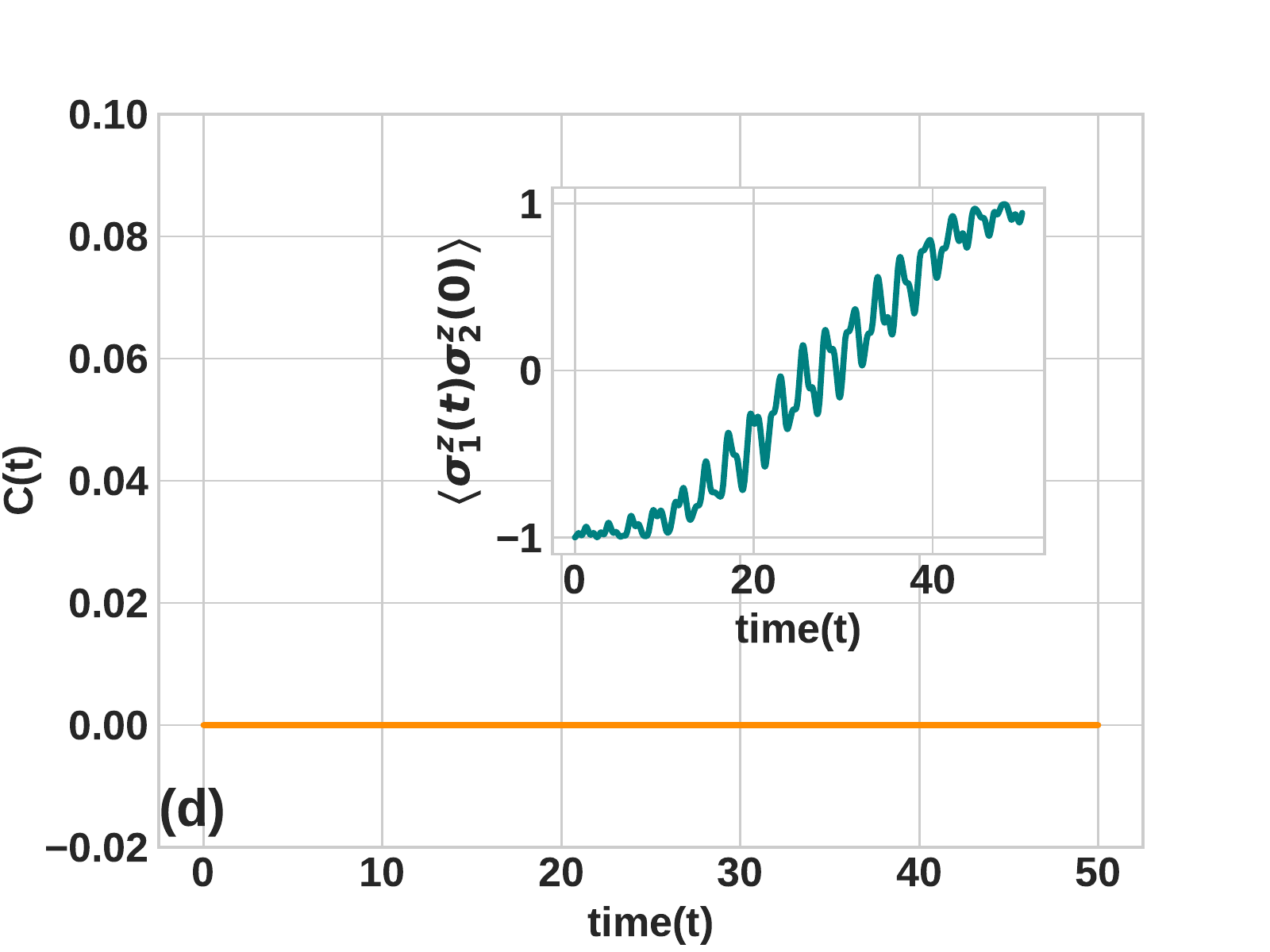}
\caption{Autonomous linear oscillators and strong connectivity regime: Position and phase space plots   in (a) and (b), and spin dynamics and OTOC in (c) and (d). Inset in (d) shows two-point time ordered correlation.  The values of the parameters are $\omega_{0}=1.5$, $\omega_{1}=1.0$, $\omega_{2}=1.5$, $F_{1}=F_{1}=0$, $\xi=0$, $\gamma=0$, $g=1$, $K=10$, $\alpha=\pi/3$. }
\label{regular1}
 \end{figure*}
The classical subsystem of the NEMS is the essence of two coupled non-autonomous nonlinear oscillators. 
Apart from the Hamiltonian part
\begin{eqnarray}\label{the Hamiltonian part}
H_0&=&\frac{1}{2}(\dot{x}_1^2+\dot{x}_2^2)+\frac{1}{2}\omega_1^2x_1^2+\frac{1}{2}\omega_2^2x_2^2   
+\frac{1}{4}\xi x_{1}^4\nonumber \\
&+&\frac{1}{4}\xi x_{2}^4+\frac{1}{2}D(x_1-x_2)^2,
\end{eqnarray}
time dependence of the oscillators $x_1(t)$, $x_2(t)$ are governed by external driving and damping terms \cite{Chotorlishvili_2011,Karabalin2009} and supplemented by quantum feedback term.
\begin{eqnarray}\label{1model}
f_1+F\cos\Omega t&=&\ddot{x}_1+\omega_1^2x_1+D(x_1-x_2)+g\langle\psi\vert\hat S^z_1\vert\psi\rangle,\nonumber\\
f_2+F\cos\Omega t&=&\ddot{x}_2+\omega_2^2x_2-D(x_1-x_2)+g\langle\psi\vert\hat S^z_2\vert\psi\rangle,\nonumber\\
\dfrac{d\vert\psi\rangle}{dt}&=&-\frac{i}{\hbar}\hat H_{s}\vert\psi\rangle,\nonumber\\
 \frac{d\langle\hat\sigma_{1,2}^j\rangle}{dt}&=&\frac{i\omega_0}{2\hbar}\langle\psi(x_1,x_2)\vert\left[\hat\sigma_{1,2}^z,\hat\sigma_{1,2}^j\right]\vert\psi(x_1,x_2)\rangle \nonumber\\
&+&\frac{ig}{\hbar}x_{1,2}(t)\langle\psi(x_1,x_2)\vert\left[\hat S_{1,2}^z,\hat\sigma_{1,2}^j\right]\vert\psi(x_1,x_2)\rangle.\nonumber\\
\end{eqnarray}
Here
\begin{eqnarray}\label{dissipation and driving part}
f_{1,2}=-2\gamma \dot{x}_{1,2}-\xi x_{1,2}^3
\end{eqnarray}
describes the effect of the nonlinear and damping terms on the dynamics  where $\gamma$ is damping constant, $\xi$ is nonlinearity constant, $F$, $\Omega$ are the amplitude and frequency of the external driving and $D$ is linear coupling term. The feedback terms $\langle\psi\vert\hat S^z_{1,2}\vert\psi\rangle$ describe the effect of the NV spin on the oscillator dynamics. We numerically solve the set of coupled equations Eq.(\ref{1model}) using Runge-Kutta Method (RK45) to integrate the wave function. The resonator is quantum if cooled down below the extremely low temperature $T<50$ nano Kelvin \cite{maroulakos2020local}. Otherwise resonator performs classical oscillations and nonlinear terms are relevant when oscillation amplitude is large. 
Due to the feedback effect, Eq.(\ref{1model}) is the essence of the coupled quantum-classical dynamics. The remarkable fact is that NV spins are not coupled to each other directly but through the nonlinear classical oscillators. Later, we will also consider the inherently quantum case where the quantum case for oscillators will be considered. 
The coupling strength between the oscillators are quantified through the connectivity 
\begin{eqnarray}\label{connectivity}
K=\frac{D}{\vert\omega_1^2-\omega_2^2\vert}.
\end{eqnarray}
In the case of weak connectivity, $K<1$ oscillators perform independent oscillations, and therefore, we do not expect arise of quantum correlations between the spins. The strong connectivity $K>1$ leads to the correlations between oscillators. Therefore, the wave function is not separable $\vert\psi(x_1,x_2,\hat\sigma_1,\hat\sigma_1,)\rangle\neq\vert\psi(x_1,\hat\sigma_1)\rangle\otimes\vert\psi(x_2,\hat\sigma_2)\rangle$. Due to non-separability, evolved spin operators are function of the position of both the oscillators 
$\hat\sigma_{1,2}(t)\equiv\hat\sigma_{1,2}(x_1(t),x_2(t))$. However, correlation between the spins $\left[\hat\sigma_{1,2}(t),\hat\sigma_{2,1}\right]\neq 0$ occur only if they exert quantum feedback on the oscillators, i.e. $x_{1,2}(t,\hat\sigma_{1,2})$,
and $\hat\sigma_{1}(x_1(t),x_2(t,\hat\sigma_{2}(t)))$.
To exclude the artefacts of different dynamical regimes, we solve Eq.(\ref{1model}) numerically for all the possible cases of interest:
\begin{enumerate}
    \item Autonomous linear system $F=0$ and $f_{1,2}=0$.
     \item Autonomous nonlinear system $F=0$, $\gamma=0$ and $\xi\neq0$.
     \item Driven linear system $F\neq0$, $\gamma\neq0$ and $\xi=0$.
     \item Driven nonlinear system $F\neq0$, $\gamma\neq0$ and $\xi\neq0$.
\end{enumerate}

We note that Hamiltonian of the spin system Eq.(\ref{two NV spins}) is coupled with the nonlinear resonator Eq.(\ref{the Hamiltonian part}), \cite{PhysRevB.79.165309,PhysRevLett.120.167401}.
We explore scrambling for the weak $K\ll 1$ and strong $K> 1$ connectivity in all these cases.  Taking into account the solution of the Schr\"odinger equation $\vert\psi(t)\rangle=\hat U(t,t_0)\vert\psi(t_0)\rangle$, where $\hat U(t,t_0)=\exp\lbrace-i\int\limits_{t_0}^td\tau \hat H_s(\tau)\rbrace$ is the evolution operator, we calculate  $\mathcal{F}=\langle\psi(t_0)\vert\hat U^{-1}\hat\sigma^z_1\hat U\hat\sigma^z_2\hat U^{-1}\sigma^z_1\hat U\hat\sigma^z_2\vert\psi(t_0)\rangle$. Hereinafter, we consider the initial spin state $\vert\psi(t_0)\rangle=\vert01\rangle$ unless specified otherwise and evaluate OTOC through  Eq.(\ref{_but_equivalent}).
\begin{figure*}
 \includegraphics[width=0.51\textwidth,height=2in]{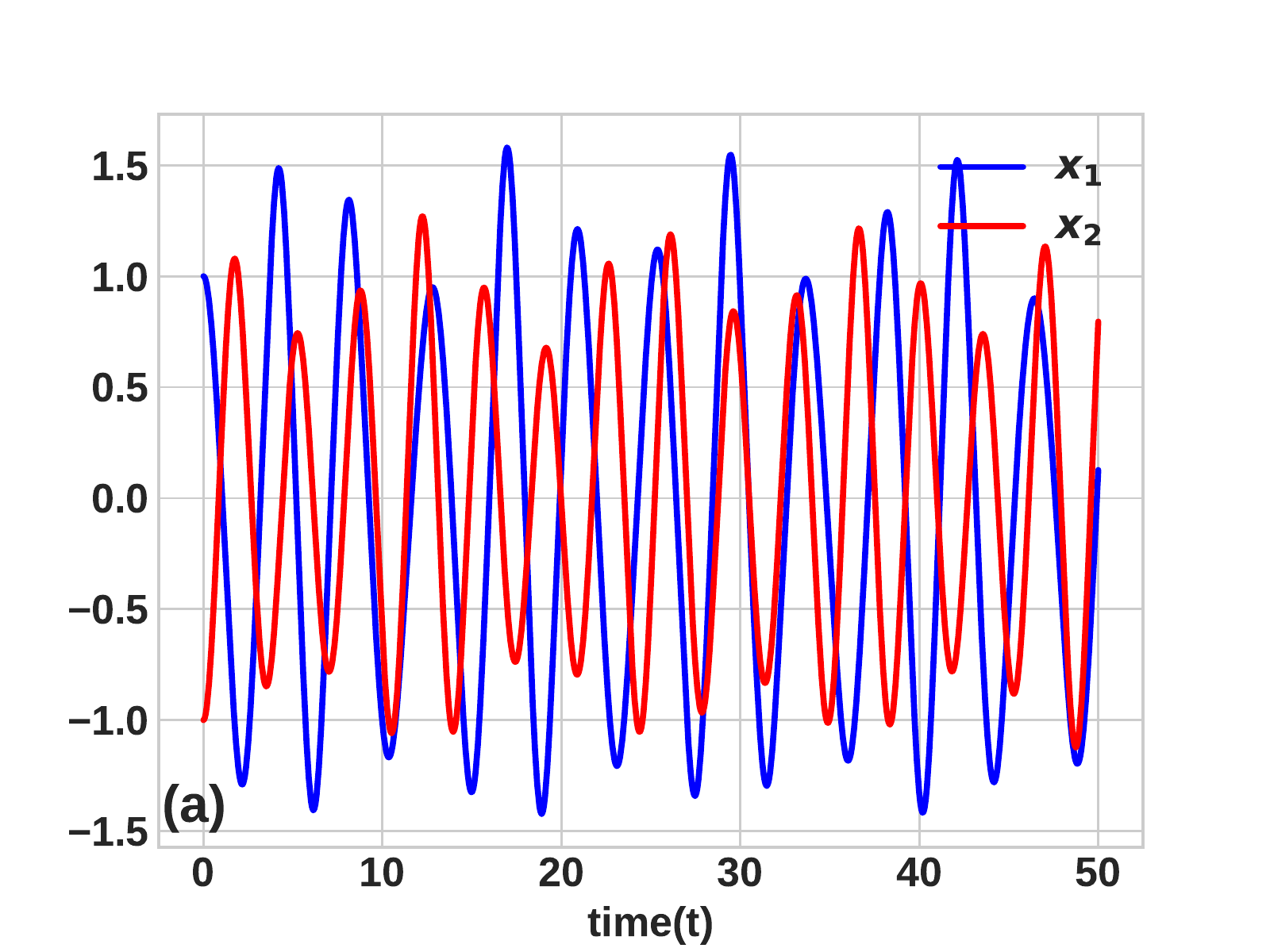}\ \includegraphics[width=0.51\textwidth,height=2in]{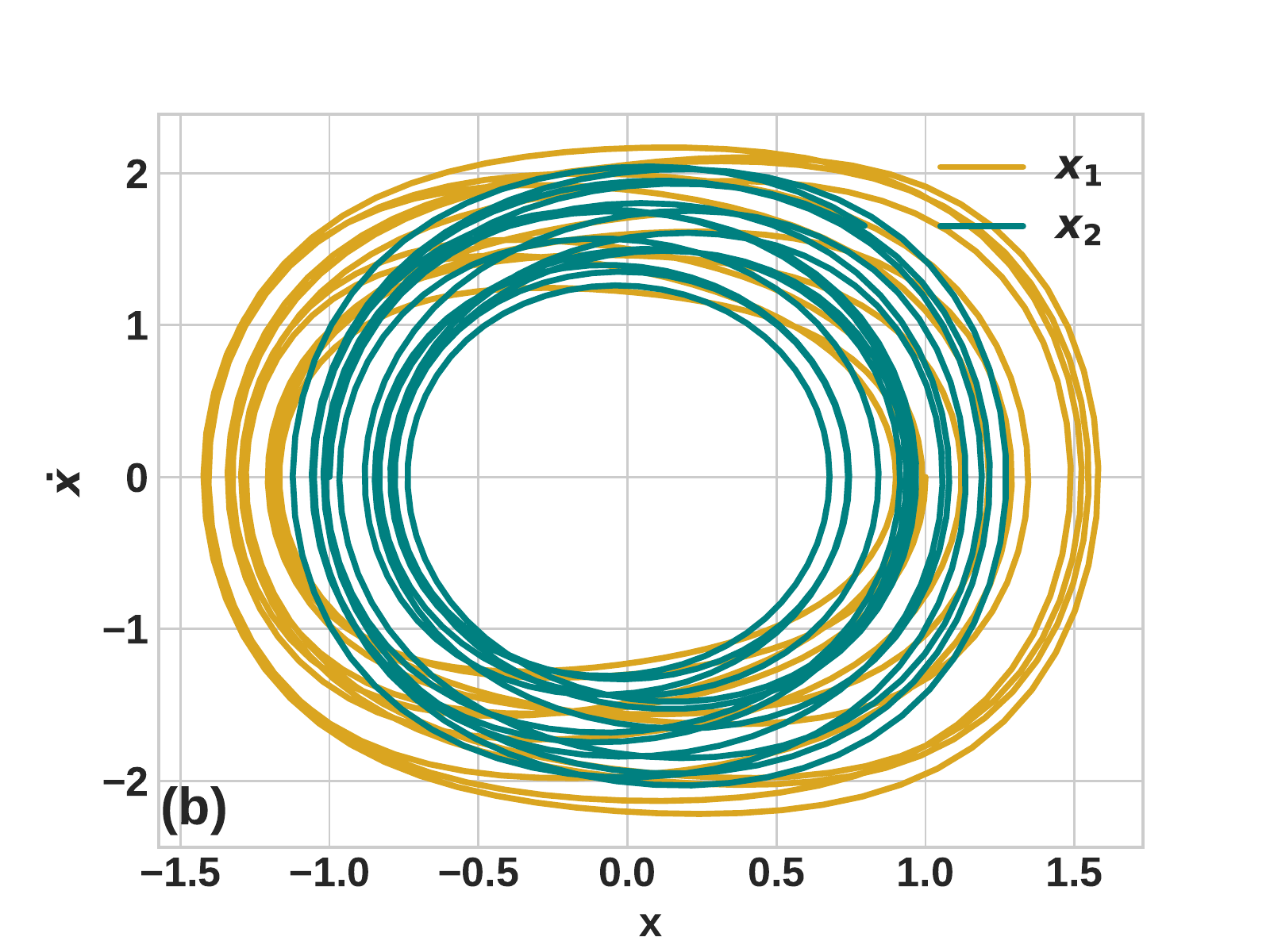}\\
 \includegraphics[width=0.51\textwidth,height=2in]{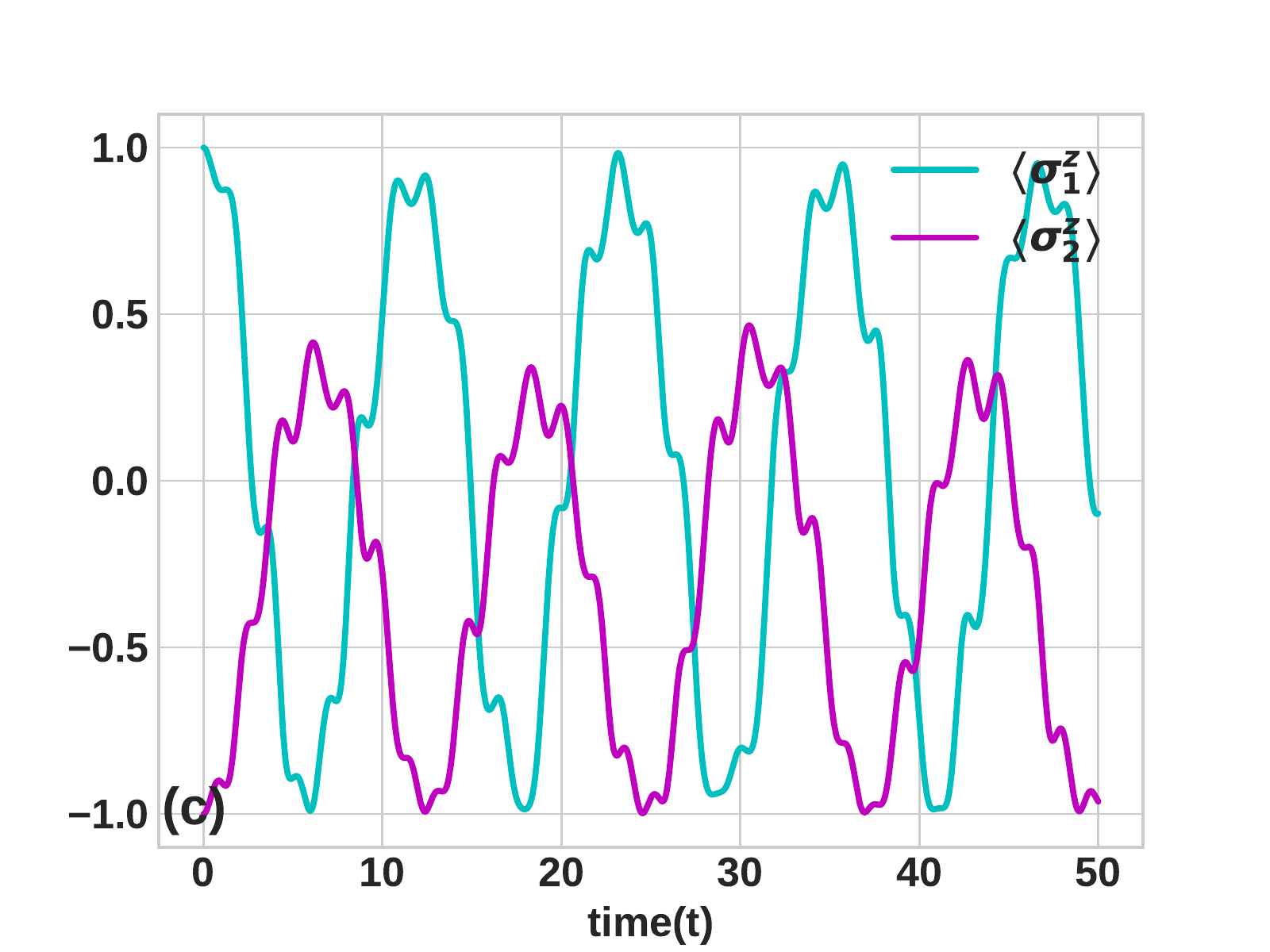}\ \includegraphics[width=0.51\textwidth,height=2in]{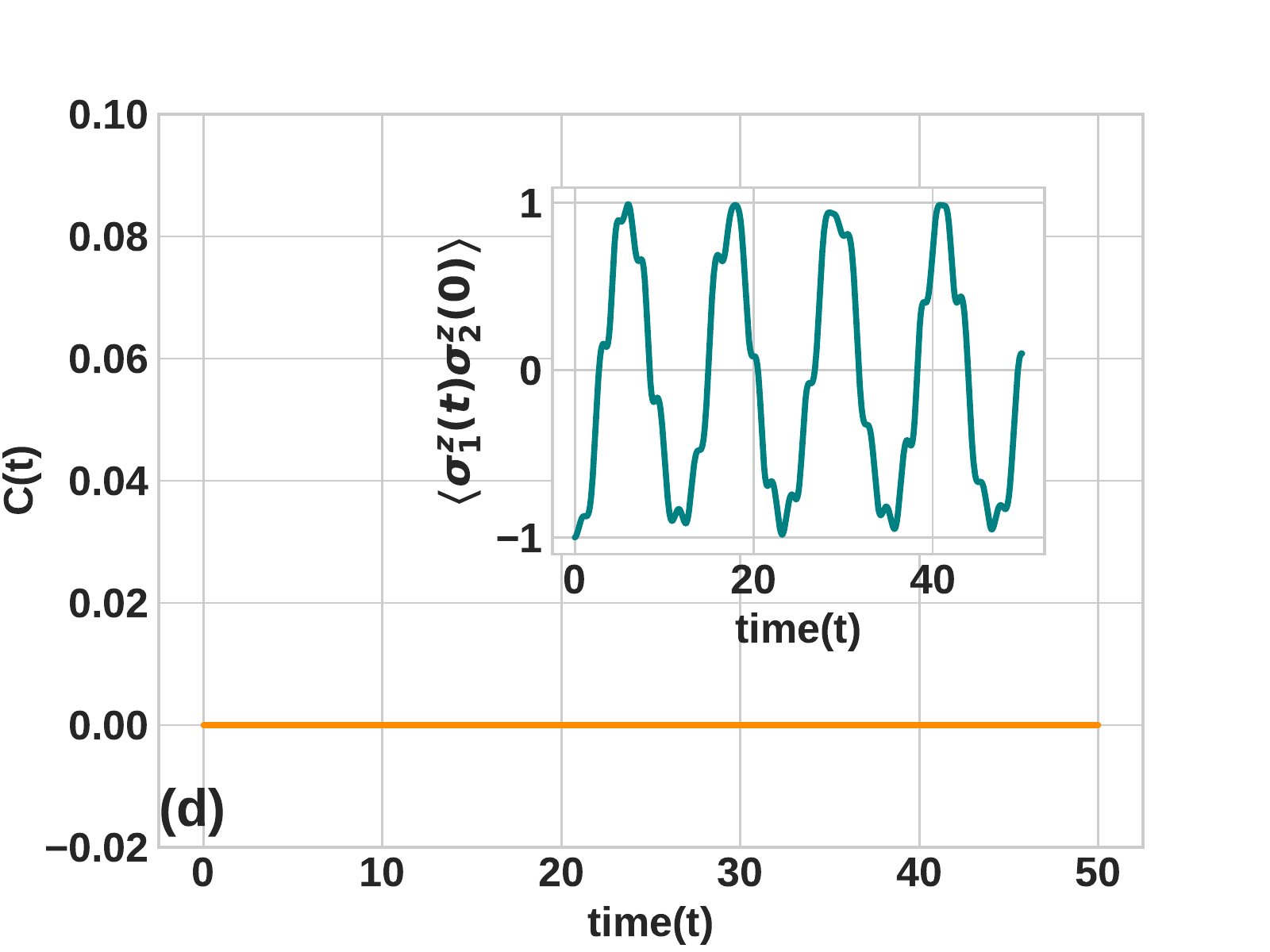}
\caption{Autonomous nonlinear oscillators and weak connectivity regime: Position and phase space plots   in (a) and (b), and spin dynamics and OTOC in (c) and (d). Inset in (d) shows two-point time ordered correlation. The values of the parameters are $\omega_{0}=1.5$, $\omega_{1}=1.0$, $\omega_{2}=1.5$, $F_{1}=F_{1}=0$, $\xi=1$, $\gamma=0$, $g=1$, $K=0.1$, $\alpha=\pi/3$. }
\label{regular2}
 \end{figure*} 
\section{Results and discussion}
\label{results}
\subsection{Analytical solution in the absence of feedback}
In the absence of quantum feedback the system admits exact analytical solution for more details see \cite{Chotorlishvili_2011}. The derivation is cumbersome and 
we present only the final result:
\begin{eqnarray}\label{cumbersome result}
&& x_{1,2}(t)=\frac{F\left(\omega_{2,1}^2-\Omega^2+2D\right)\cos\Omega t}
{4\nu_1\nu_2\sqrt{(\nu_1+\delta_1-\Omega)^2+\gamma}\sqrt{(\nu_2+\delta_2-\Omega)^2+\gamma}},\nonumber\\
&& \nu_{1,2}=\omega_1^2+\omega_2^2+2D\mp\frac{\omega_2^2-\omega_1^2}{2}\sqrt{1+K^2},\nonumber\\
&& \delta_1=\frac{3\xi}{8}\sqrt{\frac{2}{\omega_1^2+\omega_2^2}}(A_1^2+A_2^2),\nonumber\\
&& \delta_2=\frac{3\xi}{8}\sqrt{\frac{2}{\omega_1^2+\omega_2^2+4D}}(A_1^2+A_2^2),
\end{eqnarray}
where $F$, $\Omega$ are the amplitude and frequency of the external driving, $D$ is linear coupling coefficient, $\omega_{1,2}$ are frequencies of the individual resonators, $\gamma$, $\xi$ are damping and nonlinearity constant, $\delta_{1,2}$ are nonlinear corrections, and $A_{1,2}$ are the amplitudes of the induced resonator
oscillations.  We insert the solution Eq.(\ref{cumbersome result}) in $\hat H_s(x_1(t), x_2(t))$ ( given by Eq. \ref{two NV spins}) and finally get the propagated
wave function $\vert\psi(t)\rangle$=$\exp\left\lbrace -i\int\limits_{t_0}^td\tau \hat H_s(\tau)\right\rbrace \vert\psi(t_0)\rangle$, which is presented in the form:
\begin{eqnarray}\label{propagated wave function}
&&\vert\psi(t)\rangle=C_1(t)\vert 00\rangle+C_2(t)\vert 01\rangle+\nonumber\\
&&C_3(t)\vert 10\rangle+C_4(t)\vert 11\rangle.
\end{eqnarray}
Here $C_{1}(t)$, $C_{2}(t)$, $C_{3}(t)$, $C_{4}(t)$ are normalisation constant and $\vert 0\rangle,\vert 1\rangle$ is the computational basis of NV spins.
The spin dynamics reads:
\begin{eqnarray}\label{1the spin dynamics}
&& \langle\hat\sigma_1^x\rangle=2\text{Re}(C_{1}^{*}(t) C_{2}(t) +C_{3}^{*}(t) C_{4}(t)),\nonumber\\
&& \langle\hat\sigma_1^y\rangle=-2 Im (C_{1}^{*}(t) C_{2}(t) +C_{3}^{*}(t) C_{4}(t)),\nonumber\\
&& \langle\hat\sigma_1^z\rangle=(|C_1(t)|^2+|C_3(t)|^2-|C_2(t)|^2-|C_4(t)|^2),\nonumber \\
\end{eqnarray}
and
\begin{eqnarray}\label{2the spin dynamics}
&& \langle\hat\sigma_2^x\rangle=2 Re (C_{1}(t) C_{3}^{*}(t) +C_{2}(t) C_{4}^{*}(t)),\nonumber\\
&& \langle\hat\sigma_2^y\rangle=-2 Im (C_{1}(t) C_{3}^{*}(t) +C_{2}(t) C_{4}^{*}(t)),\nonumber\\
&& \langle\hat\sigma_2^z\rangle=(|C_1(t)|^2+|C_2(t)|^2-|C_3(t)|^2-|C_4(t)|^2).\nonumber \\
\end{eqnarray}
Expressions of coefficients are cumbersome and are not presented in the explicit form. 
 \begin{figure*}[tbh]
 \includegraphics[width=0.51\textwidth,height=2in]{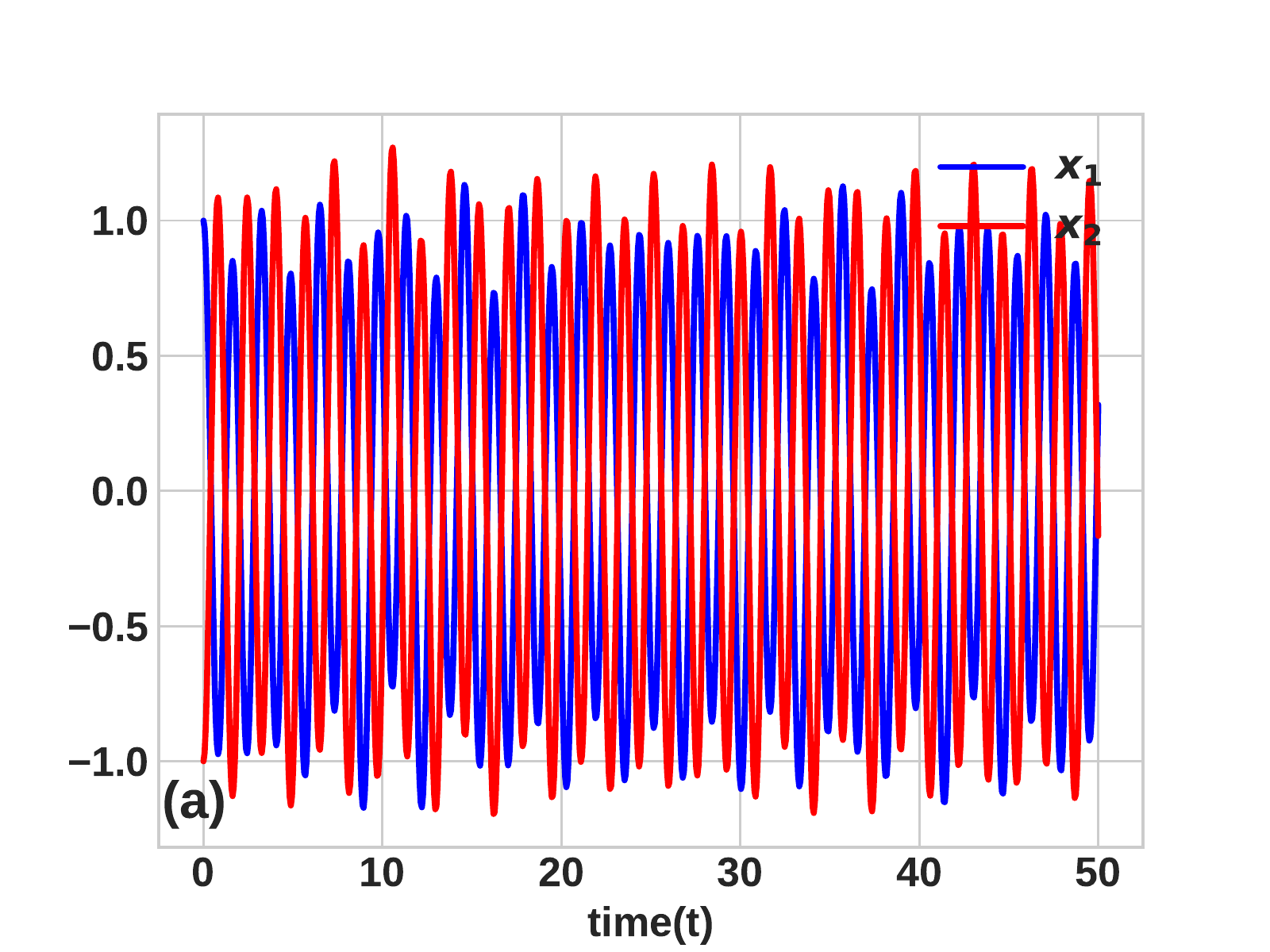}\ \includegraphics[width=0.51\textwidth,height=2in]{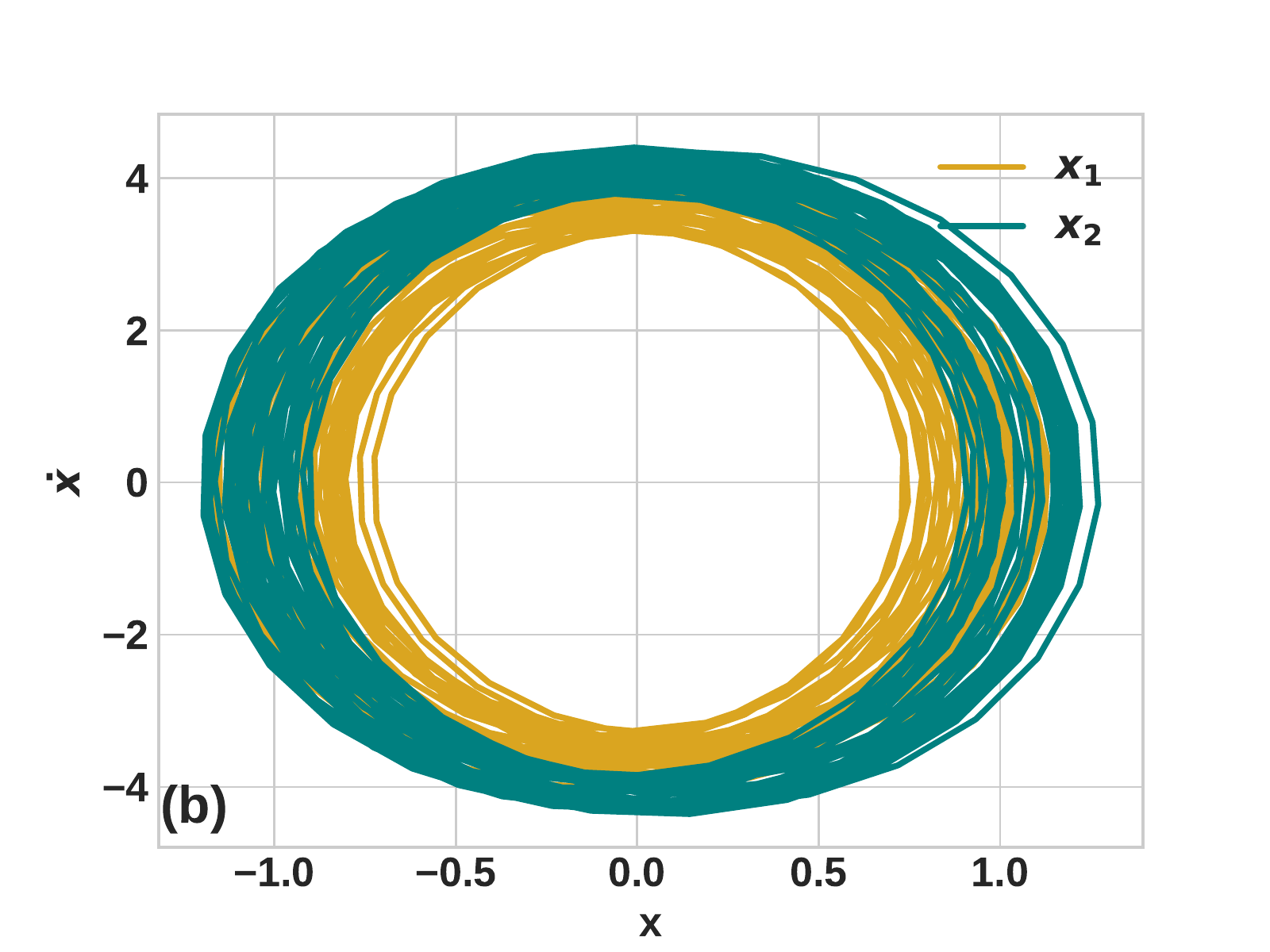}\\
 \includegraphics[width=0.51\textwidth,height=2in]{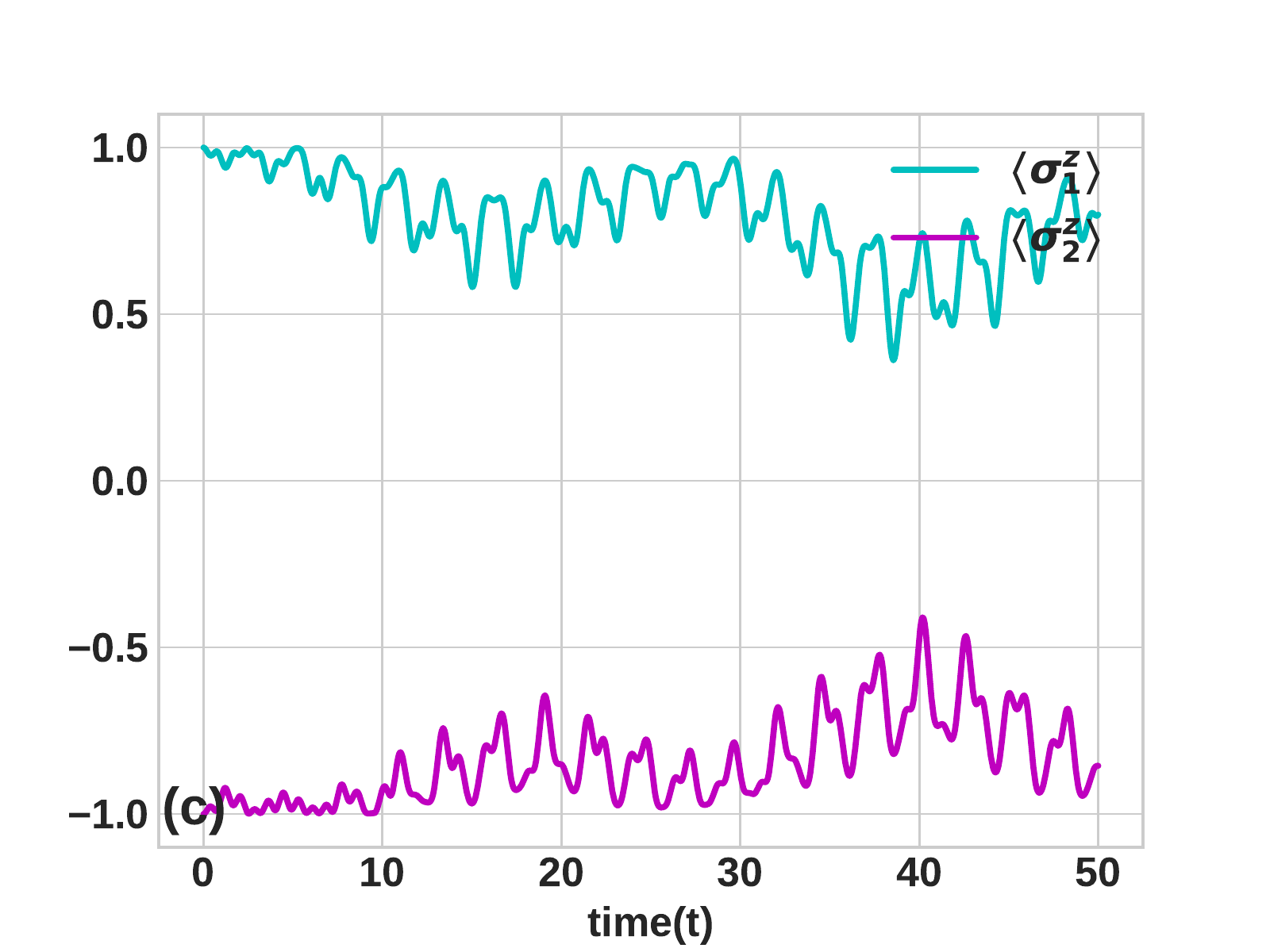}\ \includegraphics[width=0.51\textwidth,height=2in]{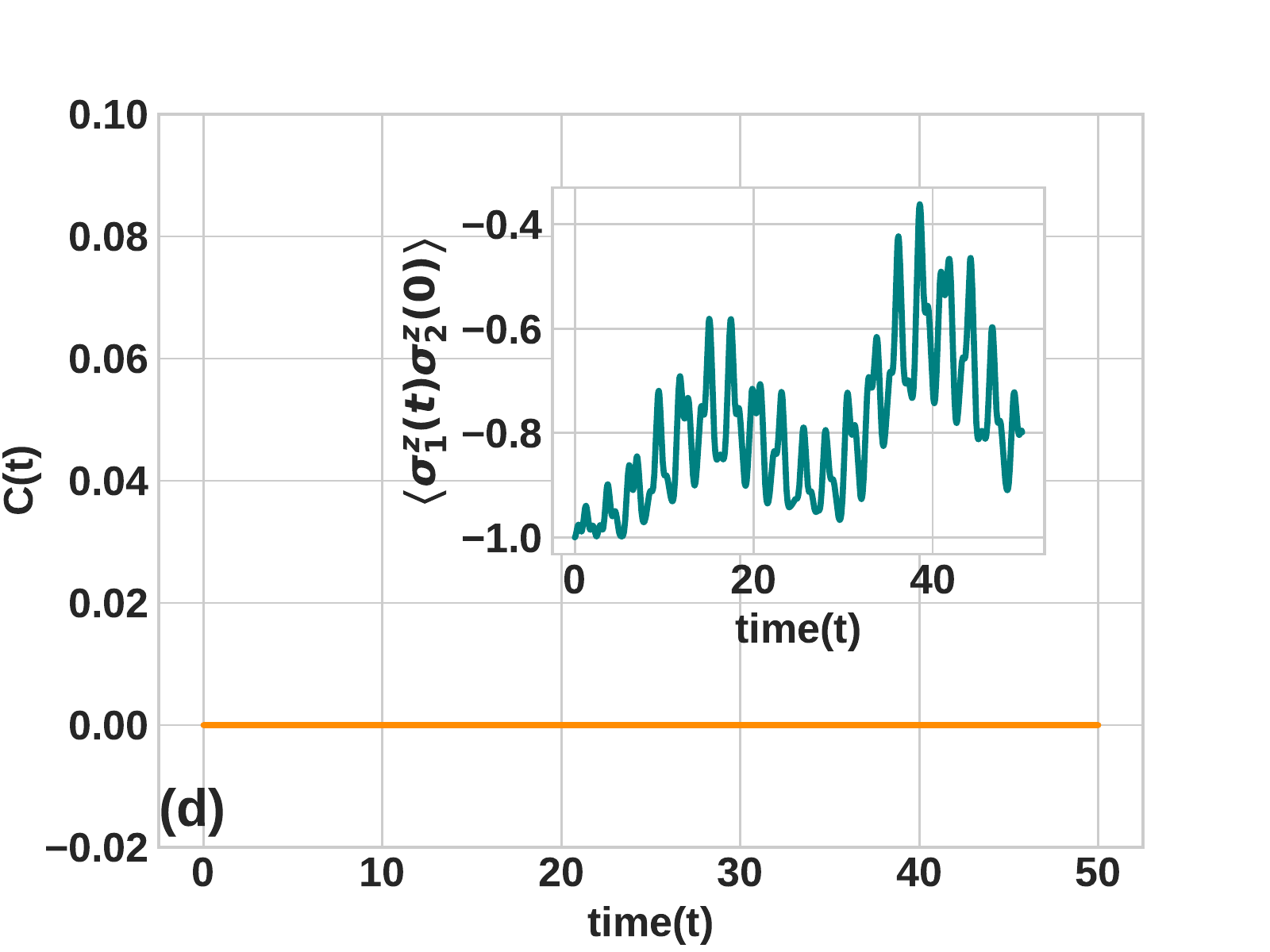}
\caption{Autonomous nonlinear oscillators and strong connectivity regime: Position and phase space plots   in (a) and (b), and spin dynamics and OTOC in (c) and (d). Inset in (d) shows two-point time ordered correlation.  The values of the parameters are $\omega_{0}=1.5$, $\omega_{1}=1.0$, $\omega_{2}=1.5$, $F_{1}=F_{1}=0$, $\xi=1$,$\gamma=0$, $g=1$, $K=10$, $\alpha=\pi/3$. }
\label{regular3}
 \end{figure*}

\subsection{Autonomous case}
In the presence of quantum feedback we use Eq.(\ref{1model}) and solve the system numerically for various cases of interest. In Fig.(\ref{regular}), we present results obtained for the linear autonomous oscillators coupled with the weak connectivity.  Dynamics of the oscillators in this case is the essence of the harmonic oscillation with a slight modulation of amplitudes. Modulation occurs due to the exchange of energy between the oscillators. In the phase portraits (Fig.\ref{regular}(b)), we see the limit cycles, closed phase trajectories of periodic motion. Fig.(\ref{regular})(c) describes the spin dynamics, the exchange of energy between the oscillator and the spin. We see periodic switching of the spin mediated by the weakly coupled to linear autonomous oscillators. The right-bottom plot Fig.(\ref{regular})(d)  shows the absence of OTOC in the system which shows a justification of the absence of quantum feedback. Two-point time ordered correlation is not zero and shows periodic switching in time ( see Inset of Fig.\ref{regular} (d)).  
  In the case of strong connectivity Fig.~\ref{regular1} dynamics of oscillators is synchronized.  The exchange of energy between the oscillators is faster leading to the trembling of the spin projection.
The spin switching is slower as shown in Fig.~\ref{regular1} (c), and OTOC (Fig.~\ref{regular1}(d)) is again zero. Even the strong connectivity regime could not inject the effects of quantum feedback in the system.  Two-point time ordered correlation shows slow switching in this case (Inset, Fig.\ref{regular1} (d)).  
 
  \begin{figure*}[h]
 \includegraphics[width=0.51\textwidth,height=2in]{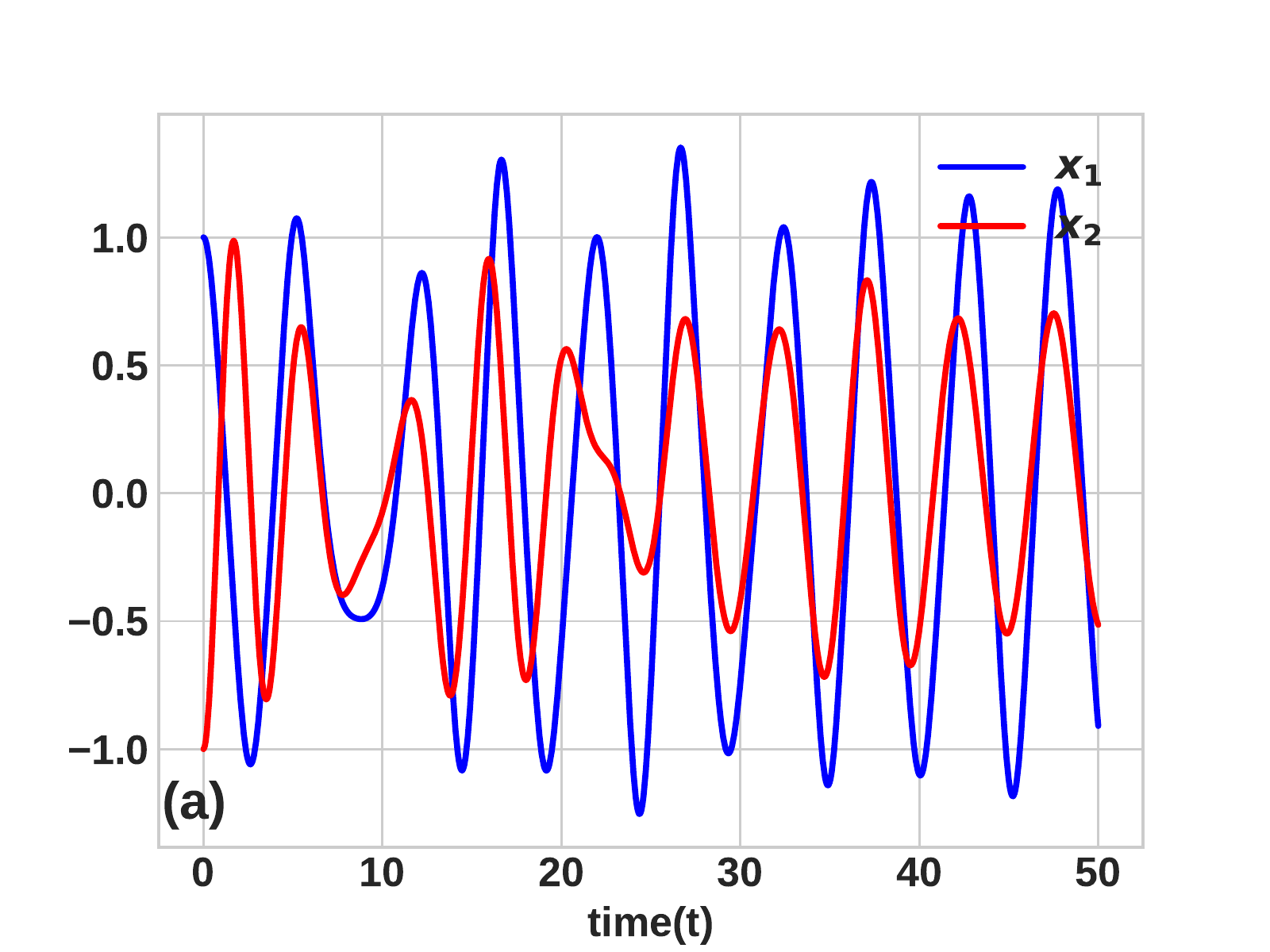}\ \includegraphics[width=0.51\textwidth,height=2in]{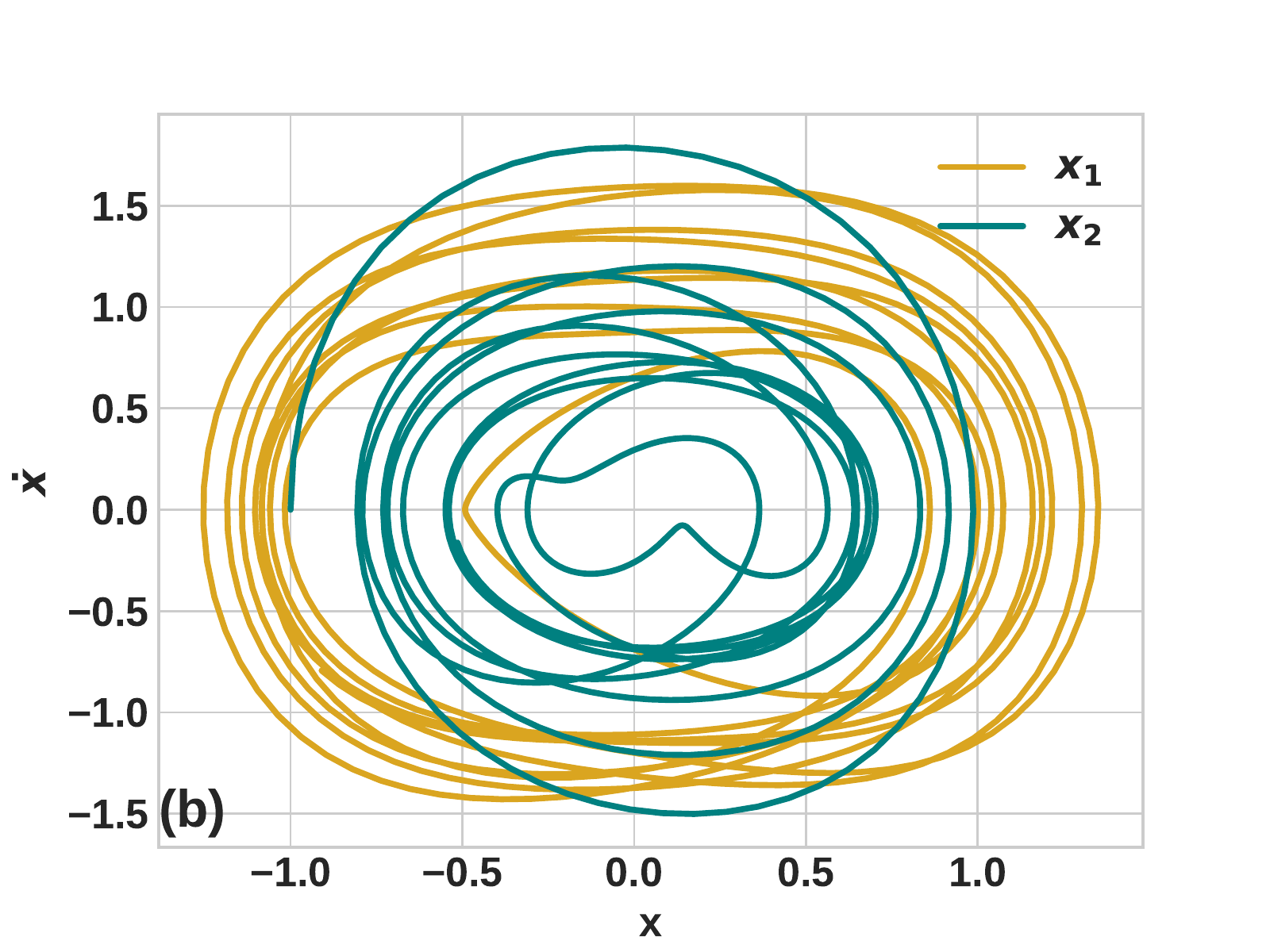}\\
 \includegraphics[width=0.51\textwidth,height=2in]{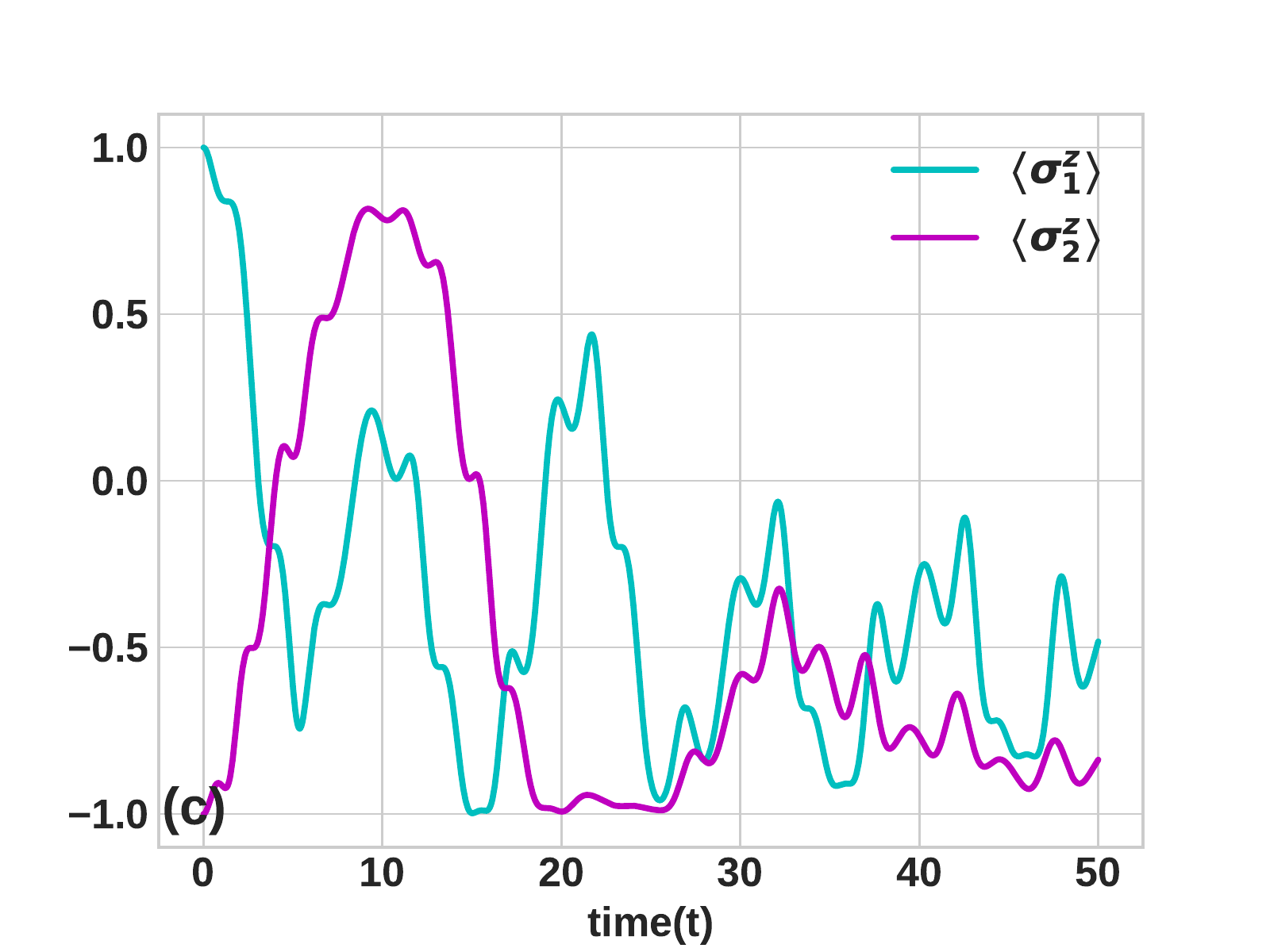}\ \includegraphics[width=0.51\textwidth,height=2in]{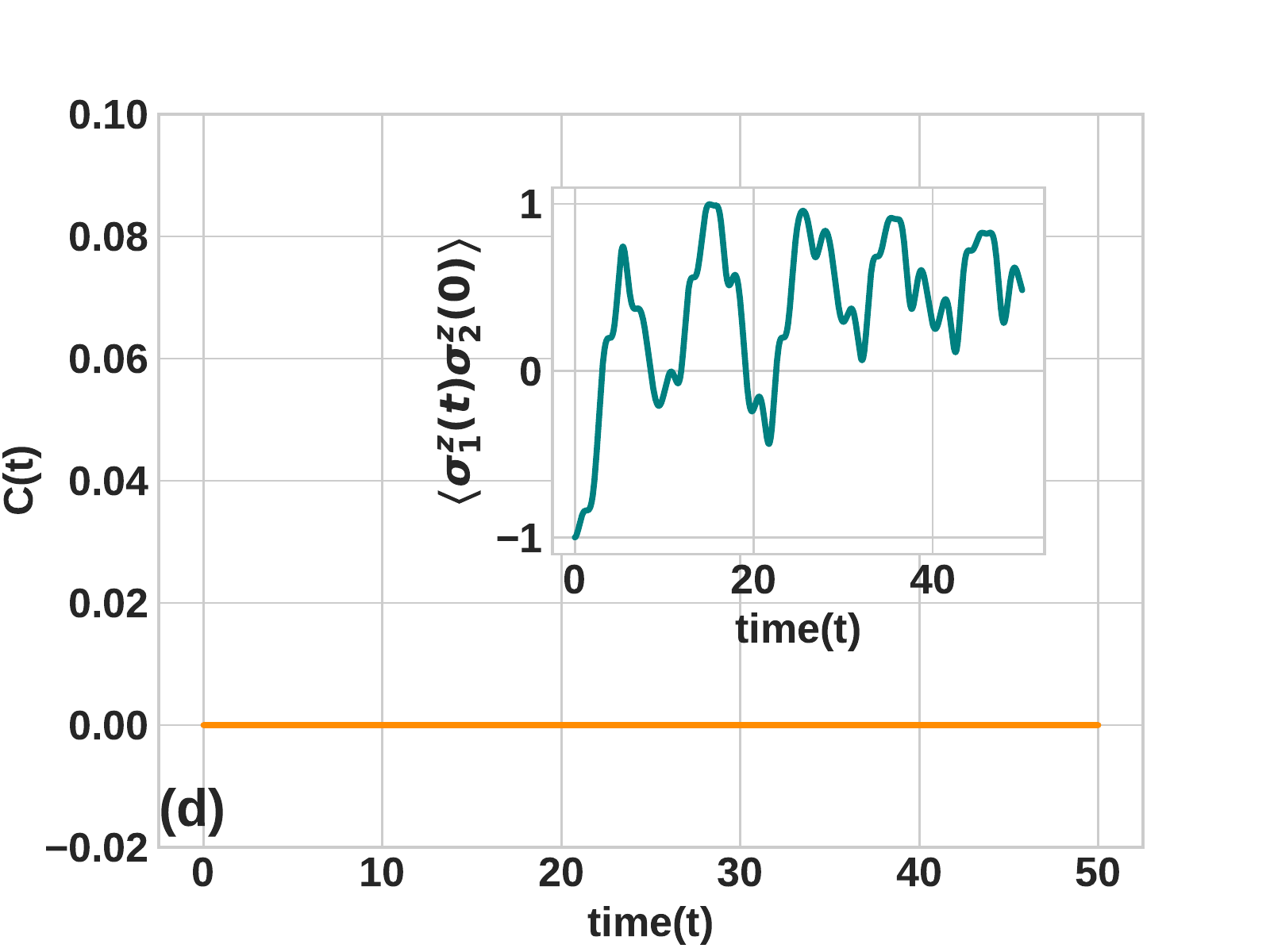}
\caption{Driven nonlinear oscillators and weak connectivity regime: Position and phase space plots   in (a) and (b), and spin dynamics and OTOC in (c) and (d). Inset in (d) shows two-point time ordered correlation.  The values of the parameters are $\omega_{0}=1.5$, $\omega_{1}=1.0$, $\omega_{2}=1.5$, $F_{1}=F_{1}=0.5$, $\xi=1$,$\gamma=0.15$, $g=1$, $K=0.1$, $\alpha=\pi/3$. }
\label{regular4}
 \end{figure*} 
Let us invoke the nonlinearity in the oscillators and explore the spin dynamics and quantum feedback in the system. In the beginning we consider the weak connectivity regime 
(see Fig.~\ref{regular2}).  As compared to the linear oscillators and the weak connectivity case,  the NV spin $\hat\sigma_2$ coupled with  the faster oscillator of position coordinate $x_2(t)$ is not switched completely. However, the NV spin $\hat\sigma_1$ switches the direction completely during the course of time. Apparently, the reason is a faster change of direction of the energy flow between the spin and the oscillator. At the halfway of switching, the energy starts to flow back to the oscillator. The quantum feedback is absent in this case also (see Fig.~\ref{regular2} (d)). We see in (Fig.\ref{regular2} (d)) inset that time ordering two point correlation shows oscillations in time with a frequency smaller than the linear case.  

Next, we explore spin dynamics and quantum feedback in the case of strongly coupled nonlinear oscillators as shown in Fig.(\ref{regular3}). We see that NV spins are frozen and perform small trembling oscillations in the vicinity of the initial values (Fig.\ref{regular3} (c)). The quantum feedback quantified in terms of OTOC (Fig.\ref{regular3} (d)) is again zero in this dynamical regime. We see in inset (Fig.\ref{regular3} (d))  that two-point time ordered correlation shows oscillations with the peak increasing in time. The nonlinearity deters the switching of the two-point time ordered correlation in the given observation.     
 \subsection{External driving}
 We know that the forced oscillations are the essence of two different evolution steps. At the early stage, the oscillation frequency coincides with the eigenfrequency of the system. However, after a specific time, the system switches to the frequency of the driving force. At first we consider the weak connectivity regime. During the transition step, we observe switching of the NV spins Fig.(\ref{regular4})(c). However, after reaching in the forced oscillation regime, the NV spins are frozen. We note that the amplitude of oscillations  does not increase resonantly due to the nonlinearity destroying the nonlinear resonance.  
 \begin{figure*}[!t]
 \includegraphics[width=0.51\textwidth,height=2in]{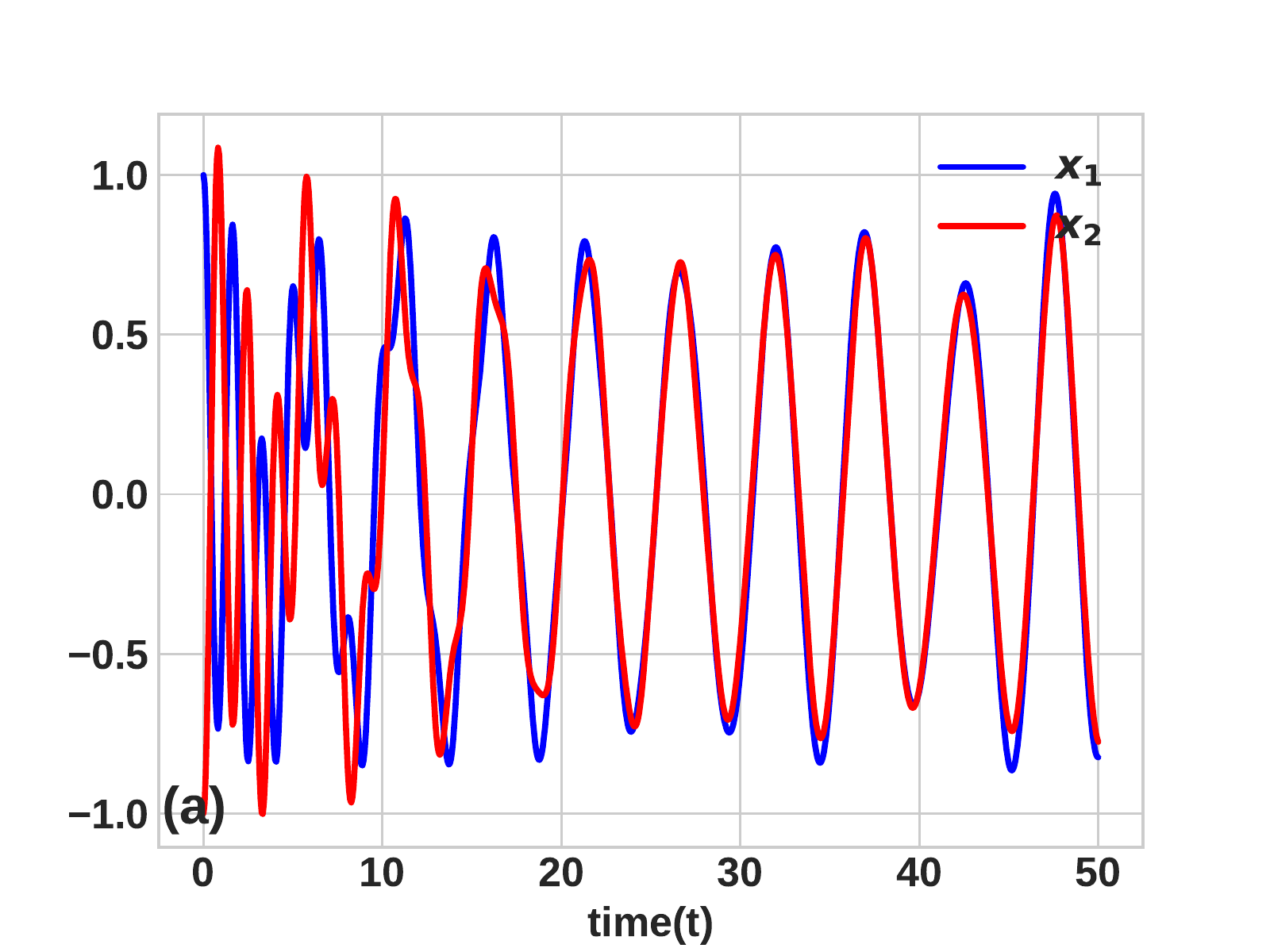}\ \includegraphics[width=0.51\textwidth,height=2in]{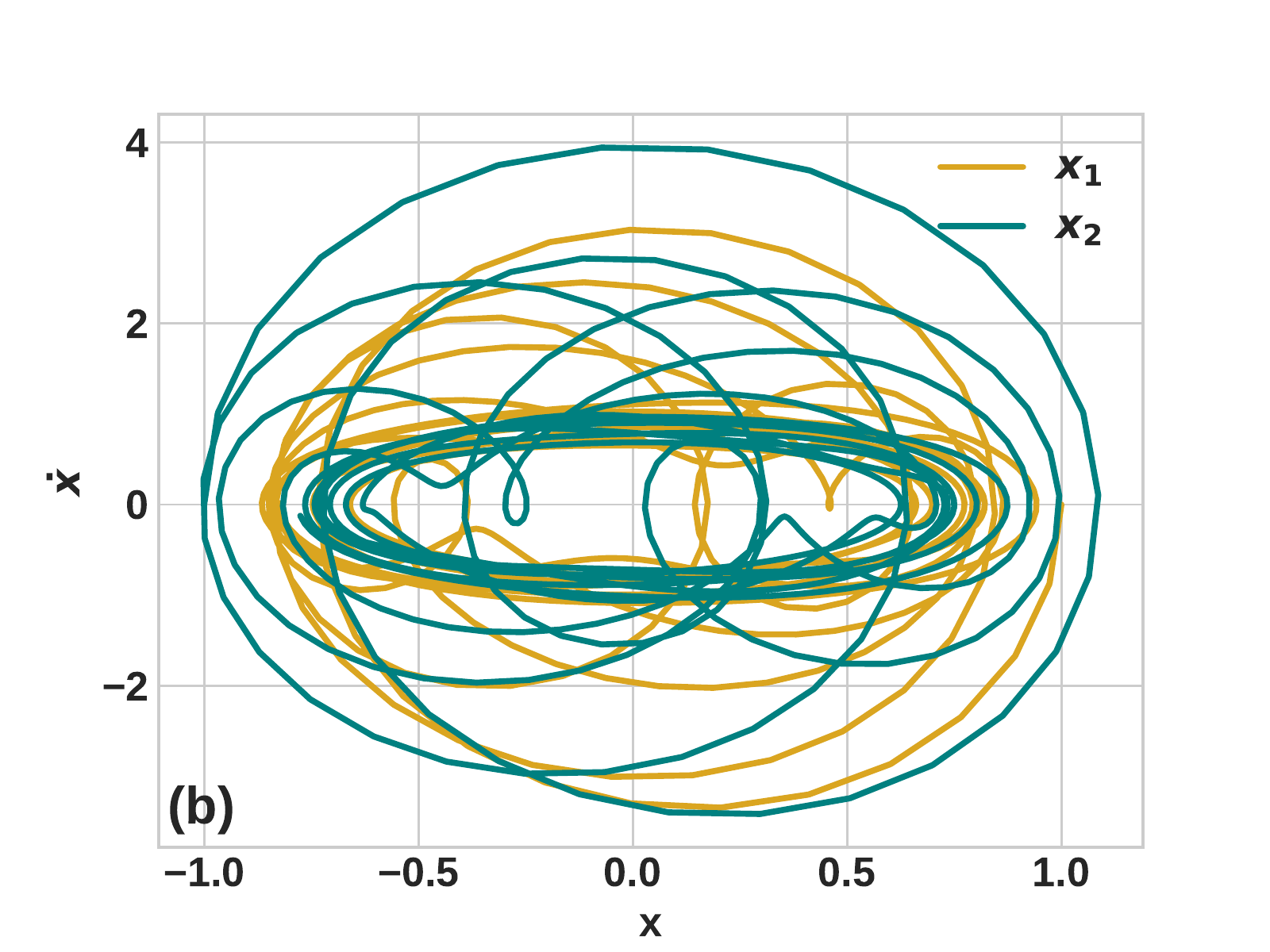}\\
 \includegraphics[width=0.51\textwidth,height=2in]{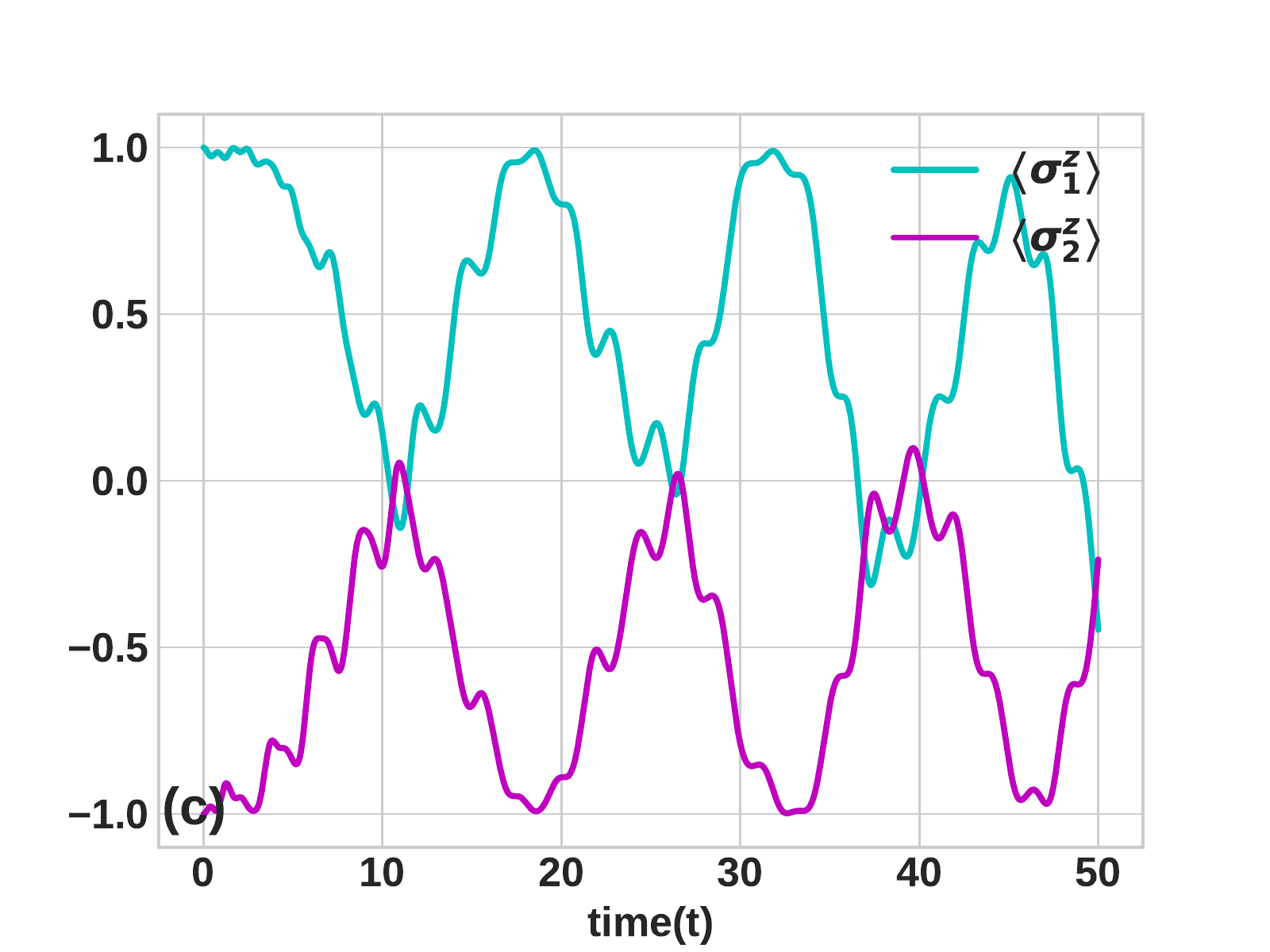}\ \includegraphics[width=0.51\textwidth,height=2in]{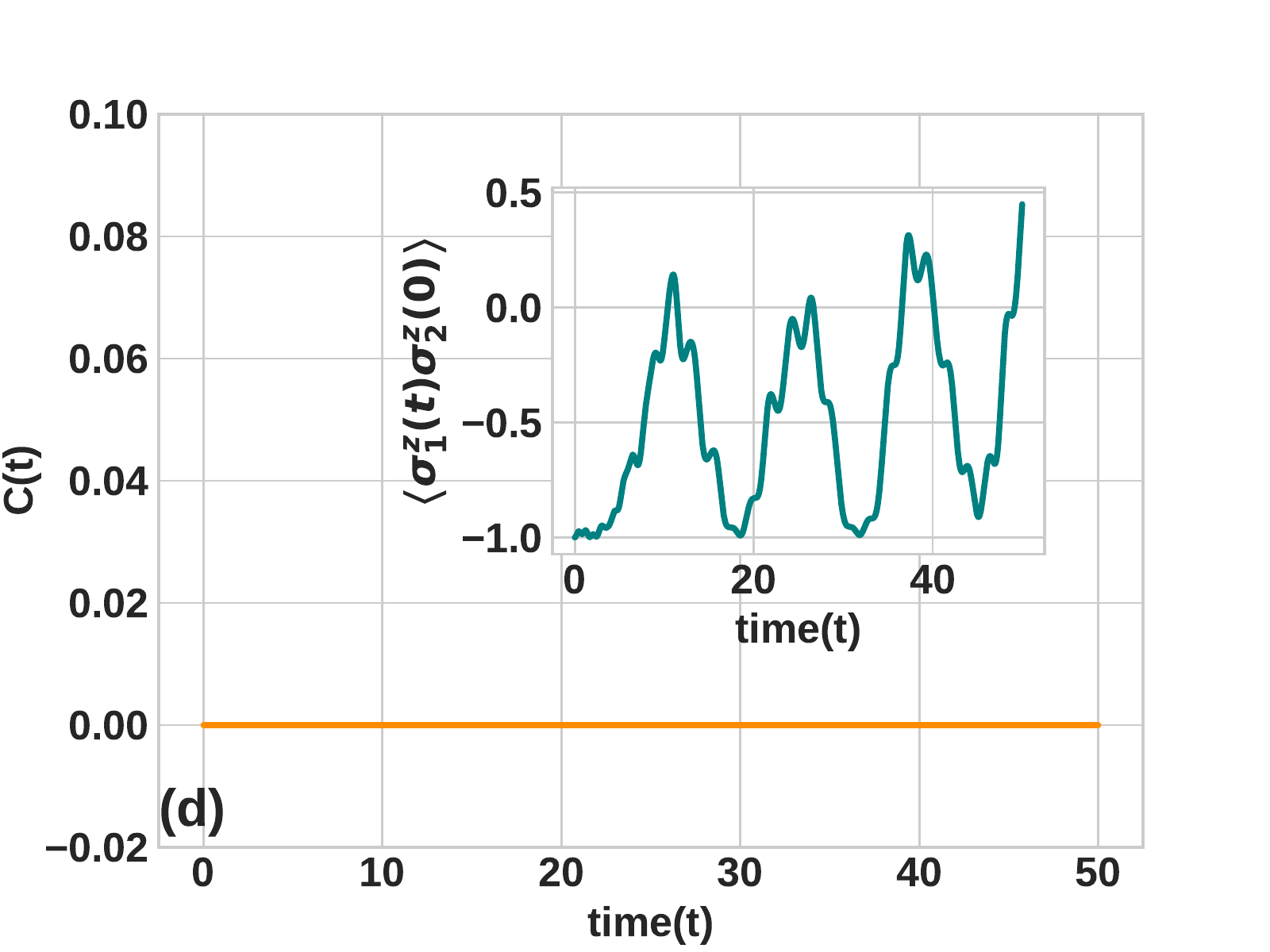}
\caption{Driven nonlinear oscillators and strong connectivity regime: Position and phase space plots   in (a) and (b), and spin dynamics and OTOC in (c) and (d). Inset in (d) shows two-point time ordered correlation.  The values of the parameters are $\omega_{0}=1.5$, $\omega_{1}=1.0$, $\omega_{2}=1.5$, $F_{1}=F_{1}=0.5$, $\xi=1$,$\gamma=0.15$, $g=1$, $K=10$, $\alpha=\pi/3$. }
\label{regular5}
 \end{figure*}
 In the case of a strong connectivity (see Fig.\ref{regular5}), oscillators are well synchronized. The NV spins and oscillator mode periodically exchange energy. In both the cases of weak and strong connectivity, OTOC and quantum feedback are absent. We see in inset of Fig.\ref{regular4} (d) and Fig.\ref{regular5} (d)  that  two-point time ordered correlation is not zero shows switching.
\\
In order to explain the absence of the quantum feedback and scrambling in the strong connectivity case, we plot the energies of classical oscillators and the quantum NV spin system. We neglect the coupling with the external driving field and dissipation process to analyse the autonomous system.  In this case, the total energy of the system comprising the NV spins and the oscillator is conserved. 
As we see Fig.(\ref{regular2x}) (a) and (b), the ratio between energies  $\langle H_{NV}\rangle/H_0=0.02$ is rather small, meaning that the energy of the oscillator is much larger than the energy of the spin system. The interaction between the oscillators and spins strongly affects NV spins and only slightly affects the oscillators. Another argument is that the relative modulation depth of energies is much larger for the quantum system:  $\delta H_0/H_0\approx 0.1$, $\delta\langle H_{NV}\rangle/\langle H_{NV}\rangle\approx 3$ and $\delta H_0+\delta\langle V\rangle+\delta\langle H_{NV}\rangle=0$. This argument becomes even stronger when external driving is switched on. The external driving field supplies the energy to the oscillators and smears out the quantum feedback effect. OTOC has also been calculated for the initial state  $\vert\psi(t_{0})\rangle=\frac{1}{\sqrt{2}}(\vert01\rangle-\vert10\rangle)$. It turned out to be zero even for the Bell states, however, the entanglement is preserved during the process.

 \begin{figure}[!t]
 \includegraphics[width=0.51\textwidth,height=2in]{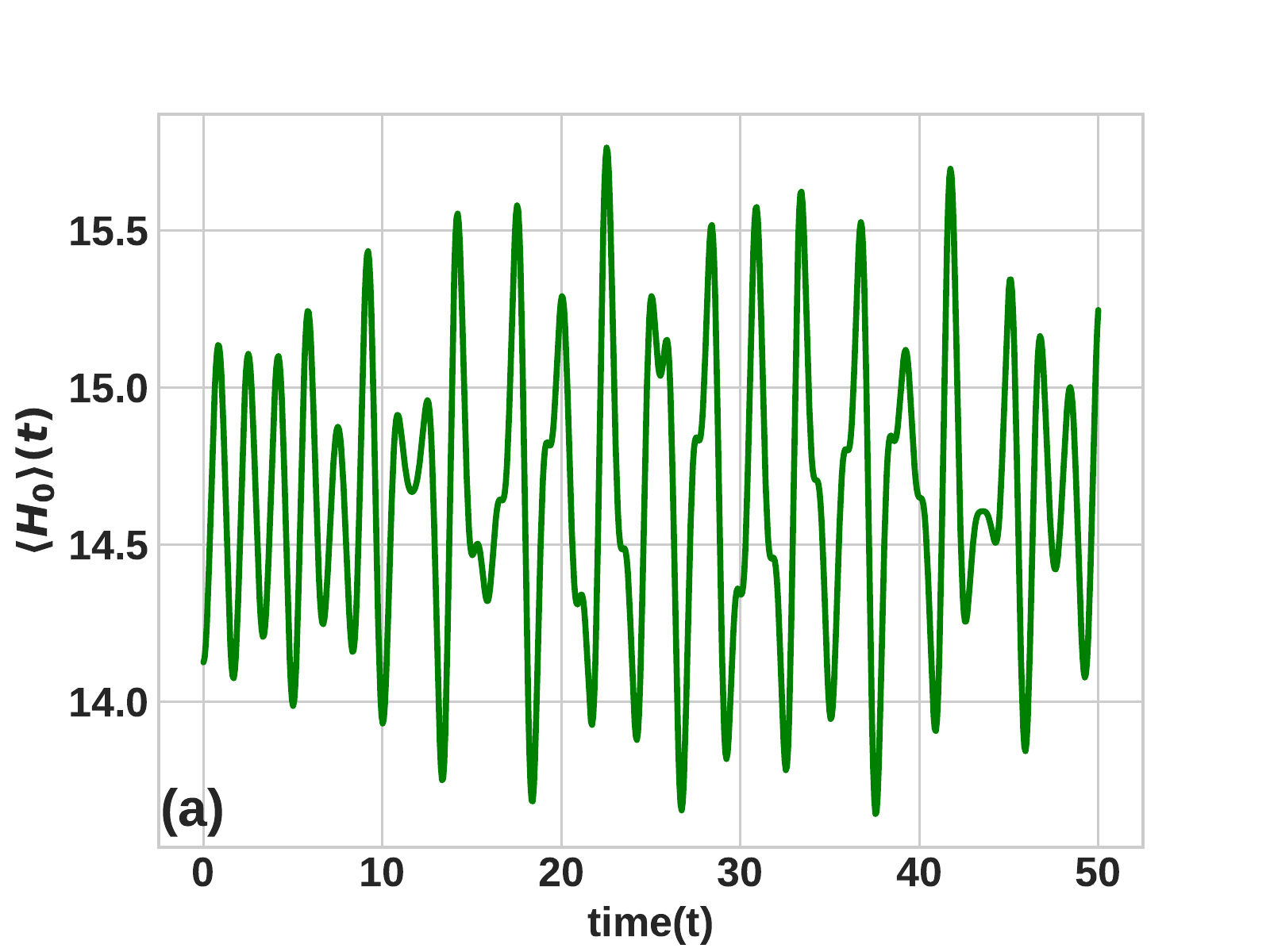}\ \includegraphics[width=0.51\textwidth,height=2in]{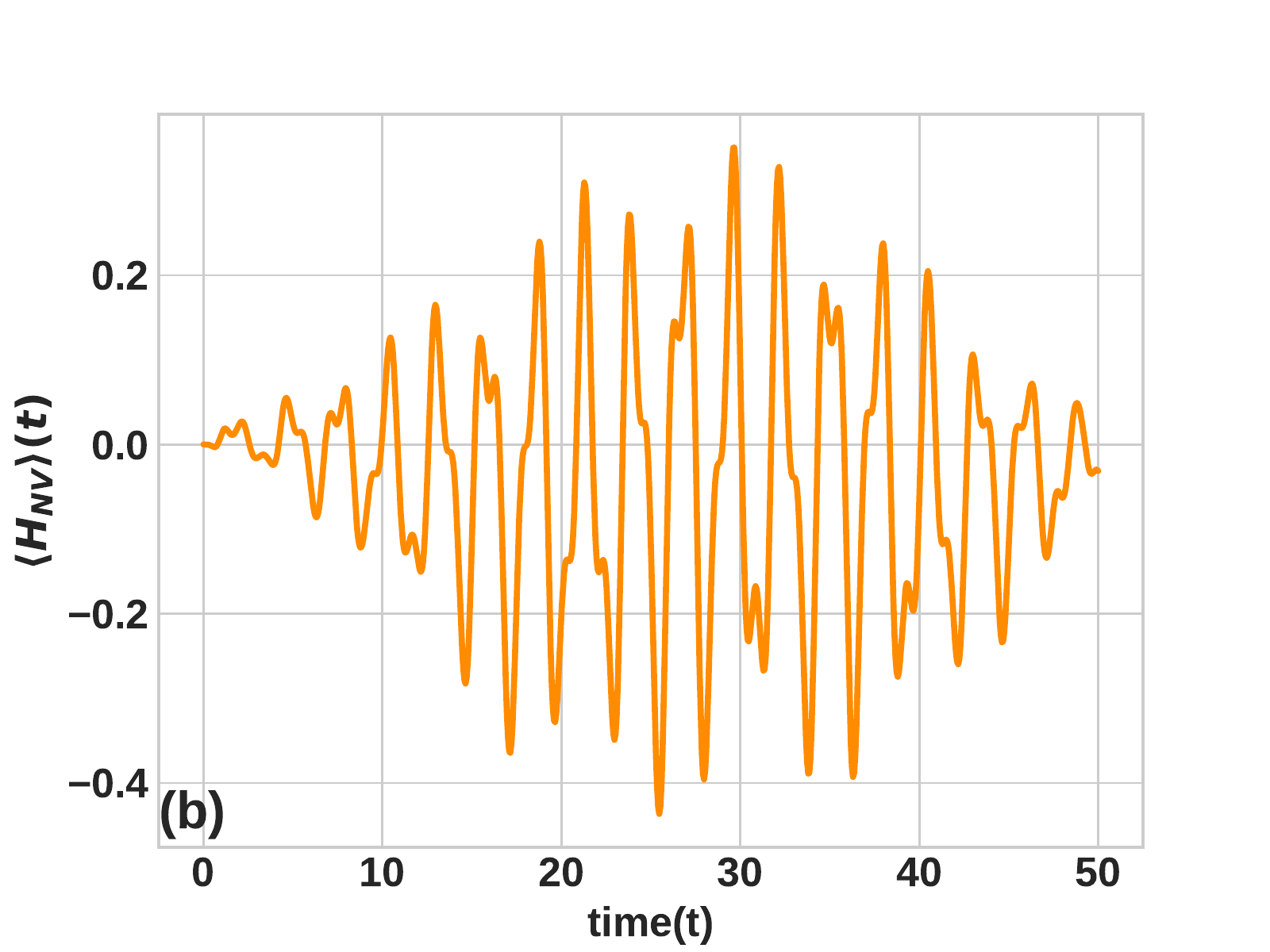}
\caption{Plot of average energy versus time for (a) the oscillator and (b) the NV spin, in the strong connectivity regime.  The values of the parameters are: $\omega_{0}=1.5$, $\omega_{1}=1.0$, $\omega_{2}=1.5$, $F_{1}=F_{1}=0$, $\xi=0$, $\gamma=0$, $g=1$, $K=10$, $\alpha=\pi/3$. The modulation depth of energies $\delta\langle H_{NV}\rangle=0.5$,  $\delta H_0=1.5$}
\label{regular2x}
 \end{figure} 
\section{Inherently quantum case: Non-zero OTOC, geometric measure of entanglement and concurrence}
\label{inherent_quantum}
In the case of strong connectivity, to some extent, coupled oscillators physically are the essence of a single effective oscillator. We show that the effective oscillator interacting with both spins indirectly couples spins.  We study three measures of quantum correlations: OTOC, concurrence, and a geometric measure of entanglement \cite{PhysRevA.99.032321,PhysRevA.68.042307,PhysRevA.77.062304}. We consider Bell's state as an initial state of the system and show that concurrence and a geometric measure of entanglement do not capture the effect of the quantum feedback. On the other hand, OTOC precisely quantifies the effect of quantum feedback. However, this effect disappears in the classical limit. Before proceed further,  once again, we specify the Hamiltonian of the system.
\begin{eqnarray}\label{spins_indirectly_coupled1}
&&\hat H=\hat H_0+g\hat V,\nonumber\\
&&\hat H_0=\omega_0\left(\hat\sigma_1^z+\hat\sigma_2^z\right)+\omega\hat a^+\hat a,\nonumber\\
&&\hat V=\lbrace\hat a^+(\hat\sigma_1^-+\hat\sigma_2^-)+\hat a(\hat\sigma_1^++\hat\sigma_2^+)\rbrace.
\end{eqnarray}
Here $\omega_0$ is the frequency of NV spin, $\omega$ is the frequency of oscillator and $g$ is NV spin-oscillator coupling constant.
For the sake of simplicity we limit the discussion of the quantum case to the linear system only. We note that inclusion of nonlinear terms leads to the dynamical Stark shift $(\omega_0+\xi\hat a^+\hat a)\hat\sigma_{1,2}$ and two-photon processes $\hat\sigma_{1,2}^+\hat a^2$, $\hat\sigma_{1,2}^-(\hat a^+)^2$, see \cite{chotorlishvili2011thermal} for more details. Nonlinear terms may affect results only quantitatively.

We follow the Fr\"ohlich method \cite{kittel1963quantum} and consider transformation of the Hamiltonian
$\tilde{H}=\exp(-\hat S)\hat H\exp(-\hat S)$, where operator $\hat S$ satisfies the condition $g\hat V+\left[\hat H_0, \hat S\right]=0$ and ensures the absence of the linear non-diagonal terms proportional to $g$ term in the transformed Hamiltonian \cite{kittel1963quantum}.  
Hamiltonian of effective interaction between two NV spins mediated by linear quantum cantilever can be derived as follows
\cite{maroulakos2020local}:
\begin{eqnarray}\label{mediated by linear quantum cantilever}
&&\hat H_{eff}=\frac{ig^2}{2}\int\limits_\infty^0d\tau\left[\hat V(\tau), V(0)\right],\nonumber\\
&&\hat V(t)=\exp(-i\hat H_0 t)\hat V\exp(i\hat H_0 t).
\end{eqnarray}
Taking into account Eq.(\ref{spins_indirectly_coupled1}) and  Eq.(\ref{mediated by linear quantum cantilever})
we deduce:
\begin{eqnarray}\label{spins_indirectly_coupled2}
&&\hat H_{eff}=\frac{g^2\left(2\hat a^+a+1\right)}{\omega_0-\omega}\times\nonumber\\
&&\left\lbrace\left(\hat\sigma_1^z+\sigma_2^z\right)+
\frac{\hat\sigma_1^+\hat\sigma_2^-+\hat\sigma_1^-\hat\sigma_2^+}{\left(2\hat a^+a+1\right)}\right\rbrace.
\end{eqnarray}
The total Hamiltonian has the form:
\begin{eqnarray}\label{spins_indirectly coupled3}
\hat H_{tot}=\hat H_0+\hat H_{eff}.
\end{eqnarray}
In what follows we replace $n=\langle\hat a^+a\rangle$ and rescale the total Hamiltonian $\hat H_{tot}/\left(2n+1\right)$.
In case of local unitary and Hermitian Pauli matrices the following expression of OTOC can be deduced from Eq. (\ref{_but_equivalent}):
\begin{eqnarray}\label{expression_OTOC}
&&C(t)=1-Re\left[\langle\hat\sigma_1^z(t)\hat\sigma_2^z\hat\sigma_1^z(t)\hat\sigma_2^z\rangle\right]
\end{eqnarray}
where time dependence is governed by total Hamiltonian as $\hat\sigma_1^z(t)=e^{-it\hat H_{tot}}\hat\sigma_1e^{it\hat H_{tot}}$.
Taking into account Eq.(\ref{spins_indirectly coupled3}) and the initial Bell's state $\vert\Phi^-\rangle=1/\sqrt{2}(\vert 01\rangle-\vert 10\rangle)$ (which is also an eigenstate of the system), for OTOC from Eq.(\ref{expression_OTOC}) we deduce:
\begin{eqnarray}\label{final OTOC}
&&C(t)= 2\sin^2\left(4\Omega_n t\right),
\end{eqnarray}
where $ \Omega_n=\frac{g^2}{(\omega_0-\omega)(2n+1)}$. 
As we see from Eq.(\ref{final OTOC}), in the initial moment of time OTOC is zero
$C(t=0)=0$ and becomes non zero for $t>0$. However in the semi-classical limit $n\rightarrow \infty$, 
$\Omega_n\rightarrow 0$ and to detect effect of the quantum feedback we need to wait rather long time
$t\approx 1/\Omega_n$, meaning that effect of the quantum feedback in rather weak (zero). 

For a two-qubit system  geometric measure of entanglement (GME) and concurrence are related to each other \cite{PhysRevA.68.042307}. To calculate concurrence we follow standard recipes:
\begin{eqnarray}\label{concurrence standard recipe}
\mathcal{C}\left[ t\right]=max\left(0, R_1-R_2-R_3-R_4\right), 
\end{eqnarray}
where $R_n(t)$ are the square roots  from the eigenvalues of the following matrix 
\begin{eqnarray}\label{eigenvalues of the R matrix}
\hat R(t)=\hat{\varrho}_R(t)\left(\hat \sigma_1^y\otimes\hat \sigma_2^y\right)\hat{\varrho}_R^*(t)\left( \hat \sigma_1^y\otimes\hat \sigma_2^y\right), 
\end{eqnarray}
$\hat{\varrho}_R(t)=Tr_{field}(\hat{\varrho}(t))$ is the reduced density matrix $\hat{\varrho}(t)=e^{-it\hat H_{tot}}\hat{\varrho}(0)e^{it\hat H_{tot}}$, $\hat{\varrho}(0)=\vert\Phi^-\rangle\langle\Phi^-\vert$ and $\hat\sigma_{1,2}^y$ are Pauli matrices acting on the first and the second spins. The GME is given by GME$=(1-\sqrt{1-\mathcal{C}\left[ t\right]})/2$ \cite{PhysRevA.68.042307}. For the Hamiltonian Eq.(\ref{spins_indirectly coupled3}) and the initial Bell's state $\vert\Phi^-\rangle=1/\sqrt{2}(\vert 01\rangle-\vert 10\rangle)$, we calculate
$\mathcal{C}\left[ t\right]=1$ and GME$=0.5$.
Thus in the case of a particular initial state, 
the concurrence and GME are constant, and through them, we cannot quantify quantum feedback exerted by spins on the oscillator. On the other hand, the quantum feedback if described by OTOC,  will be nonzero if present (as in quantum case) and vanish if absent (as in the classical limit).
\begin{figure}[!t]
\centering
 \includegraphics[width=0.5\textwidth,height=2.2in]{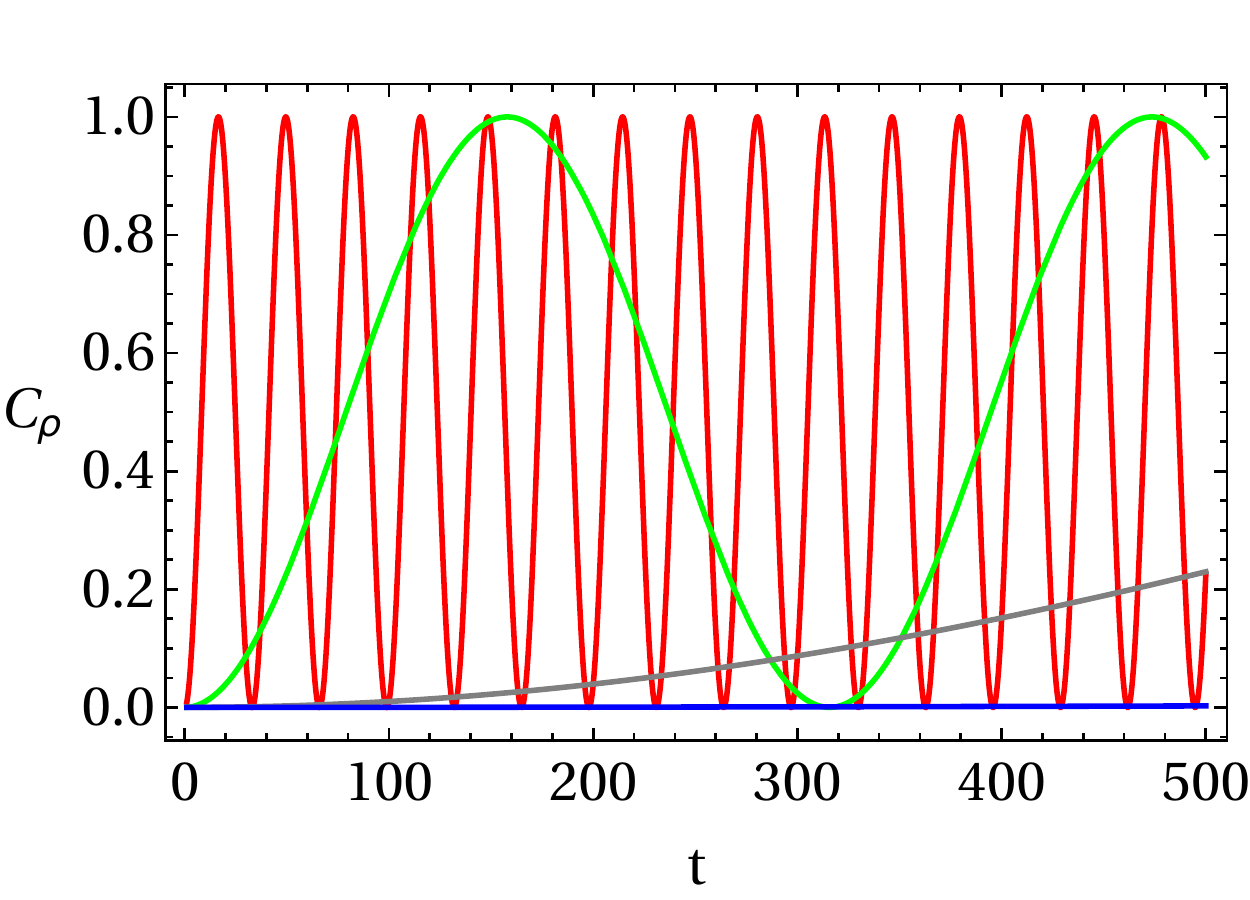}
\caption{Plot of thermally averaged OTOC vs time at fixed  Temperature(T). The parameter used are $g=1$, $\omega_{0}=3$, $\omega=2$, $T=100$, $n=10$(\rm{red}), $n=100$(\rm{Green}), $n=1000$(\rm{Gray}), $n=10000$(\rm{blue}). For very large $n$ the oscillations die out. }
\label{ThermalOTOC}
 \end{figure} 
We can also check the finite temperature case. At a finite temperature, in the equilibrium state, the density matrix in the basis of Hamiltonian is given as 
\begin{eqnarray}\label{thermal_density}
&&\hat\varrho=Z^{-1}\begin{pmatrix}
e^{-2\beta (\Omega_0+\omega_{0R})}& 0 & 0 & 0\\
0 & e^{- \beta \Omega_{n}} &0&0\\
0  & 0&e^{ \beta \Omega_{n}}&0\\
0 & 0 & 0 & e^{2\beta (\Omega_0+\omega_{0R})}
\end{pmatrix},\nonumber\\
&& Z=2\cosh{\beta 2(\Omega_0+\omega_{0R})}+2\cosh{\beta \Omega_{n}},
\end{eqnarray}
where we introduced the notations $\Omega_0=\frac{g^2}{(\omega_0-\omega)}$, $\omega_{0R}=\frac{\omega_{0}}{2n+1}$ and $\beta=1/T$ is the
inverse temperature.
 Thermally averaged OTOC $ C_\rho=1-Re(Tr\lbrace \hat\varrho\hat\sigma_1^z(t)\hat\sigma_2^z\hat\sigma_1^z(t)\hat\sigma_2^z\rbrace)$ is calculated (for details see Appendix \ref{appendix1}) as:
\begin{eqnarray}\label{Thermal}
 C_\rho=1-\frac{\cosh 2\beta (\Omega_0+\omega_{0R})+\cos4\Omega_n t\cosh{\beta\Omega_n}}{\cosh{2\beta (\Omega_0+\omega_{0R})}+\cosh{\beta \Omega_{n}}},\nonumber\\
\end{eqnarray}
and the thermal concurrence 
$\mathcal{C}_\rho[t]$ is calculated using Eq. (\ref{concurrence standard recipe}) as
\begin{eqnarray}
\mathcal{C}_\rho[t]=2\times{\rm max}\Big(0,\frac{\vert\sinh{\beta\Omega_n}\vert-1}{\big(2\cosh{2\beta (\Omega_0+\omega_{0R})}+2\cosh{\beta \Omega_{n}}\big)}\Big).\nonumber\\
\end{eqnarray} 
Thermally averaged OTOC as a function of time is plotted in Fig.(\ref{ThermalOTOC}) for different $n$.
At the finite temperatures,  in the semi-classical limit $n\rightarrow \infty$, 
we have $\Omega_n\rightarrow 0$,  $\omega_{0R}\rightarrow 0$, $C_\rho=0$ and $\mathcal{C}_\rho[t]=0$. On the other hand, at a zero temperature dynamical effect, i.e., the quantum feedback is captured only by OTOC. Both at finite or zero temperature OTOC is not zero if spin exerts feedback on the oscillator, and it becomes zero when $n\rightarrow \infty$.  
\section{Conclusions}
\label{conclusion}
Out of time correlation function is widely used as a quantitative measure of spreading quantum correlations.  In the present work, we propose an experimentally feasible model of the nanomechanical system for which OTOC can be exploited as a quantifier of quantum feedback. In particular, we consider two NV spins coupled with two different oscillators.  Oscillators are coupled to each other directly, and NV spins are not.  Therefore, any quantum correlation between the NV spins may arise only due to the quantum feedback exerted by NV spins on the corresponding oscillators.  The OTOC operator between the NV spins is the essence of the quantifier of the quantum feedback. To exclude the artefacts of a particular type of oscillations, we considered different dynamical regimes: linear vs. nonlinear, free, and forced oscillations and showed that the OTOC and quantum feedback are zero in all the cases. We also considered quantum oscillator and quantum spins case where the indirect coupling between the spins is invoked by a quantum harmonic oscillator. We have shown nonzero OTOC, GME and concurrence in this case. In a classical limit of the oscillator, the OTOC vanishes. Thus, we conclude that entanglement between two NV centers can spread only if NV centers are connected through the quantum channel. In the semi-classical and classical channel limit, entanglement decays to zero.

\appendix
\section{Appendix: Calculation of thermally averaged OTOC $C_\rho$}
\label{appendix1}
From Eq.(\ref{spins_indirectly coupled3}) after re-scaling the total Hamiltonian will be written as:
\begin{eqnarray}
H_{tot}=(\omega_{0R}+\Omega_0)(\sigma_{1}^z+ \sigma_{2}^z)+\Omega_n(\sigma_{1}^+ \sigma_{2}^- + \sigma_{1}^- \sigma_{2}^+) 
\end{eqnarray}
or in the matrix form, in the standard basis,
\begin{eqnarray}
\begin{pmatrix}
2(\Omega_0+\omega_{0R}) & 0 & 0 & 0\\
0 &0 &\Omega_{n} &0\\
0  & \Omega_{n}&0&0\\
0 & 0 & 0 & -2(\Omega_0+\omega_{0R})
\end{pmatrix},
\end{eqnarray}
where $n=\langle a^{\dagger}a\rangle$, $\omega_{0R}=\frac{\omega_{0}}{2n+1}$,
and $\Omega_0=\frac{g^2 }{\omega_{0}-\omega}$, and $\Omega_{n}=\frac{g^2 }{(\omega_{0}-\omega)(2n+1)}$.
The eigenstates of the above Hamiltonian are
\begin{eqnarray}
\vert \phi_{1}\rangle=\vert 0,0\rangle,\\
\vert \phi_{2}\rangle=\frac{1}{\sqrt{2}}(\vert 1,0\rangle+\vert 0,1\rangle),\\
\vert \phi_{3}\rangle=\frac{1}{\sqrt{2}}(\vert 1,0\rangle-\vert 0,1\rangle),\\
\vert \phi_{4}\rangle=\vert 1,1\rangle,
\end{eqnarray}
with corresponding eigenvalues:
\begin{eqnarray}
E_{1}=2(\Omega_0+\omega_{0R})\\
E_{2}=\Omega_{n}\\
E_{3}=-\Omega_{n}\\
E_{4}=-2(\Omega_0+\omega_{0R}).
\end{eqnarray}
At a finite temperature, in the equilibrium state the density matrix $\hat\varrho=Z^{-1}e^{-\beta H_{tot}}$  of the system in the diagonal basis of the Hamiltonian is
\begin{eqnarray}
\hat\varrho&=&Z^{-1}\big(e^{-\beta E_{1}}\vert\phi_{1}\rangle\langle\phi_{1}\vert+e^{-\beta E_{2}}\vert\phi_{2}\rangle\langle\phi_{2}\vert+e^{-\beta E_{3}}\vert\phi_{3}\rangle\langle\phi_{3}\vert\nonumber\\
&+&e^{-\beta E_{4}}\vert\phi_{4}\rangle\langle\phi_{4}\vert\big)
\end{eqnarray}
\begin{eqnarray}\label{thermal_density}
&&\hat\varrho=Z^{-1}\begin{pmatrix}
e^{-2\beta (\Omega_0+\omega_{0R})}& 0 & 0 & 0\\
0 & e^{- \beta \Omega_{n}} &0&0\\
0  & 0&e^{ \beta \Omega_{n}}&0\\
0 & 0 & 0 & e^{2\beta (\Omega_0+\omega_{0R})}
\end{pmatrix},\nonumber\\
&& Z=2\cosh{\beta 2(\Omega_0+\omega_{0R})}+2\cosh{\beta \Omega_{n}}.
\end{eqnarray}
Pauli operators $\sigma_{1}^{z}$ and $\sigma_{2}^{z}$ in the diagonal basis pf Hamiltonian are written as:
\begin{eqnarray}
\sigma_{1}^{z}=\vert\phi_{1}\rangle\langle\phi_{1}\vert+\vert\phi_{3}\rangle\langle\phi_{2}\vert+\vert\phi_{2}\rangle\langle\phi_{3}\vert-\vert\phi_{4}\rangle\langle\phi_{4}\vert
\end{eqnarray}
\begin{eqnarray}
\sigma_{2}^{z}=\vert\phi_{1}\rangle\langle\phi_{1}\vert-\vert\phi_{3}\rangle\langle\phi_{2}\vert-\vert\phi_{2}\rangle\langle\phi_{3}\vert-\vert\phi_{4}\rangle\langle\phi_{4}\vert
\end{eqnarray}
Also the time evolution operators $\exp{(-i H_{tot} t)}$ in the diagonal basis can be given as:
\begin{eqnarray}
\exp{(-i H_{tot} t)}&=&e^{- i E_{1}t}\vert\phi_{1}\rangle\langle\phi_{1}\vert+e^{- i E_{2}t}\vert\phi_{2}\rangle\langle\phi_{2}\vert\nonumber\\
&+&e^{- i E_{3}t}\vert\phi_{3}\rangle\langle\phi_{3}\vert+e^{= i E_{4}t}\vert\phi_{4}\rangle\langle\phi_{4}\vert,
\end{eqnarray}
We calculate $\sigma_{1}^{z}(t)$  as
\begin{eqnarray}
\sigma_{1}^{z}(t)&=&e^{iH_{tot}t}\sigma_{1}^{z}e^{-iH_{tot}t}=\vert\phi_{1}\rangle\langle\phi_{1}\vert+e^{-2i\Omega_{n}t}\vert\phi_{3}\rangle\langle\phi_{2}\vert\nonumber\\ &+&e^{2i\Omega_{n}t}\vert\phi_{2}\rangle\langle\phi_{3}\vert-\vert\phi_{4}\rangle\langle\phi_{4}\vert.
\end{eqnarray}
By successive application of operators in the sequence $\sigma_{1}^{z}(t)\sigma_{2}^{z}\sigma_{1}^{z}(t)\sigma_{2}^{z}$ we get:
\begin{eqnarray}
\sigma_{1}^{z}(t)\sigma_{2}^{z}\sigma_{1}^{z}(t)\sigma_{2}^{z}&=&\big(\vert\phi_{1}\rangle\langle\phi_{1}\vert-e^{4i\Omega_{n}t}\vert\phi_{2}\rangle\langle\phi_{2}\vert\nonumber\\
&-&e^{-4i\Omega_{n}t}\vert\phi_{3}\rangle\langle\phi_{3}\vert+\vert\phi_{4}\rangle\langle\phi_{4}\vert\big).
\end{eqnarray}
We can calculate $\rho \sigma_{1}^{z}(t)\sigma_{2}^{z}\sigma_{1}^{z}(t)\sigma_{2}^z $ as
\begin{eqnarray}
\rho\sigma_{1}^{z}(t)\sigma_{2}^{z}\sigma_{1}^{z}(t)\sigma_{2}^{z}&=&Z^{-1}\big(e^{-2\beta(\Omega_0+\omega_{0R})}\vert\phi_{1}\rangle\langle\phi_{1}\vert\nonumber\\&-&e^{-\beta \Omega_n}e^{4i\Omega_{n}t}\vert\phi_{2}\rangle\langle\phi_{2}\vert\nonumber \\
&-&e^{\beta\Omega_n}e^{-4i\Omega_{n}t}\vert\phi_{3}\rangle\langle\phi_{3}\vert\nonumber\\&+&e^{2\beta(\Omega_0+\omega_{0R})}\vert\phi_{4}\rangle\langle\phi_{4}\vert\big).
\end{eqnarray}
Further, we calculate the thermally averaged OTOC $C_\rho(t)=1-Re(Tr \{\rho \sigma_{1}^{z}(t)\sigma_{2}^{z}\sigma_{1}^{z}(t)\sigma_{2}^z\})$ as
\begin{eqnarray}
 C_\rho(t)=1-\frac{2\cosh 2\beta (\Omega_0+\omega_{0R})+2\cos4\Omega_n t\cosh{\beta\Omega_n}}{Z},\nonumber\\
\end{eqnarray}
\vspace{0.5cm}
{\bf Author contribution statement} \\
\noindent
AKS, LC and SKM conceived of the presented idea and developed the theory and performed analysis. AKS, KS and VV performed numerical calculations. All authors discussed the results and contributed to the final manuscript.
 \bibliographystyle{unsrt}
 \bibliography{bibliography}
%

%

\end{document}